\documentclass[aip,cha,amsmath,amssymb,reprint]{revtex4-2}
\usepackage{indentfirst}
\usepackage{braket}
\usepackage{makeidx}
\usepackage{slashed}
\usepackage{graphicx}
\usepackage{xcolor}
\usepackage{footmisc}
\usepackage[normalem]{ulem}
\usepackage{bm}				% bold math symbols
\usepackage{amsmath}
\usepackage{amsmath}
\usepackage{amssymb}
\usepackage{latexsym}
\usepackage{amsfonts}
\usepackage{graphicx}
\usepackage{tikz}
\usetikzlibrary{arrows.meta}
\usetikzlibrary{decorations.pathmorphing,backgrounds,patterns}
\usetikzlibrary{shapes}
\usetikzlibrary{calc}
\usepackage{pgfplots} 
\usepackage{pgfplotstable}
\usepgfplotslibrary{external}
\tikzset{>=stealth}
\tikzset{external/optimize=false}
\pgfplotsset{compat=newest}
\usetikzlibrary{pgfplots.groupplots}
\usetikzlibrary{calc}
\usetikzlibrary{3d}
\usetikzlibrary{positioning}
\usepgfplotslibrary{units}
\pgfplotsset{unit code/.code={\si{#1}}}
\usepgfplotslibrary{patchplots}
\tikzset{>=stealth}
\usepackage{dcolumn}
\usepackage[utf8]{inputenc}
\usepackage[T1]{fontenc}
\definecolor{mDarkTeal}{HTML}{23373b}
\definecolor{mLightBrown}{HTML}{EB811B}
\definecolor{myred}{RGB}{228,26,28}
\definecolor{myblue}{RGB}{55,126,184}
\definecolor{myorange}{RGB}{225,127,0}
\definecolor{mygreen}{RGB}{77,175,74}
\definecolor{mylila}{RGB}{152,78,163}
\definecolor{mybrown}{RGB}{153,76,0}
\definecolor{mygray}{RGB}{153,153,153}
\definecolor{darkred}{rgb}{0.8,0,0}
\definecolor{mydarkgreen}{RGB}{0,102,0}
\definecolor{mydarkbrown}{RGB}{102,52,0}
\definecolor{Orange}{RGB}{235,129,27}
\definecolor{Green}{RGB}{35,55,59}
\definecolor{mycolor}{rgb}{0.0,0.42,0.24}
\renewcommand{\ket}[1]{\vert{#1}\rangle} 
\renewcommand{\bra}[1]{\langle{#1}\vert} 
 
\newcommand{\op}[2]{\ket{#1}\!\bra{#2}}
\newcommand{\be}{\begin{equation}}
\newcommand{\ee}{\end{equation}}

\newcommand{\cor}[1]{\left[ #1 \right]}
\newcommand{\acor}[1]{\left\{ #1 \right\}}

\newcommand{\bc}{\begin{center}}
\newcommand{\ec}{\end{center}}

\newcommand{\ben}{\begin{eqnarray}}
\newcommand{\een}{\end{eqnarray}}

\renewcommand{\vec}[1]{\boldsymbol{#1}}

\newcommand{\cH}{{\cal H}}

\newcommand{\cL}{{\cal L}}

\newcommand{\cB}{{\cal B}}

\newcommand{\Tr}{\textrm{Tr}}

\begin{document}
\title{Degenerated Liouvillians and Steady-State Reduced Density Matrices}
\author{Juzar Thingna}
\affiliation{Center for Theoretical Physics of Complex Systems, Institute for Basic Science (IBS), Daejeon 34126, Republic of Korea.}
\affiliation{Basic Science Program, University of Science and Technology, Daejeon 34113, Republic of Korea.}
\author{Daniel Manzano}
\email{manzano@onsager.ugr.es}
\affiliation{Universidad de Granada, Departamento de Electromagnetismo y F\'isica de la Materia and Instituto Carlos I de F\'isica Te\'orica y Computacional, Granada 18071, Spain.}
%{\mailto{jythingna@ibs.re.kr}; \mailto{manzano@onsager.ugr.es}}

\begin{abstract}
Symmetries in an open quantum system lead to degenerated Liouvillian that physically implies the existence of multiple steady states. In such cases, obtaining the initial condition independent stead states is highly nontrivial since any linear combination of the \emph{true} asymptotic states, which may not necessarily be a density matrix, is also a valid asymptote for the Liouvillian. Thus, in this work we consider different approaches to obtain the \emph{true} steady states of a degenerated Liouvillian. In the ideal scenario, when the open system symmetry operators are known we show how these can be used to obtain the invariant subspaces of the Liouvillian and hence the steady states. We then discuss two other approaches that do not require any knowledge of the symmetry operators. These could be a powerful tool to deal with quantum many-body complex open systems. The first approach which is based on Gramm-Schmidt orthonormalization of density matrices allows us to obtain \emph{all} the steady states, whereas the second one based on large deviations allows us to obtain the non-degenerated maximum and minimum current-carrying states. We discuss our method with the help of an open para-Benzene ring and examine interesting scenarios such as the dynamical restoration of Hamiltonian symmetries in the long-time limit and apply the method to study the eigenspacing statistics of the nonequilibrium steady state.
\end{abstract}
\maketitle
%---------------------introduction---------------

\begin{quotation}
In 1976 Gorini, Kossakowski, Sudarshan, and Lindblad (GKSL)~\cite{gorini:jmp76,lindblad:cmp76} independently proposed a completely positive trace preserving master equation that governs the dynamics of a generic quantum system. Since then the equation has been a hallmark in the study of dissipative open quantum systems and has been used in a wider variety of applications. In recent years, due to the experimental advancements, engineering the bath properties and system-bath interaction has become possible. One immediate consequence is the existence of multiple steady states. In such cases, the dissipative Liouvillian becomes degenerated, having more than one invariant subspace. In general, finding the nonequilibrium steady states (NESS) is highly nontrivial and in this work we outline three methods to address this issue. Each method has its own benefits and drawbacks. Using a \emph{para}-Benzene ring as a open quantum system we elucidate the methods and find the existence of decoherence free subspaces or even dynamical restoration of Hamiltonian symmetries in the long time limit. Lastly, since our approach allows us to obtain the NESSs for a degenerated Liouvillian we use it to study the statistics of the ratio of consecutive eigenspacing $r$ of the NESS which shows $P(r)\rightarrow 0$ as $r\rightarrow 0$.
\end{quotation}

\section{Introduction}

Quantum master equations are an essential tool to study dissipative systems and have been applied to a wide variety of model systems in quantum optics~\cite{olmos:prl12,manzano:sr16,Hanarxiv20}, thermodynamics~\cite{thingnapre16,chiara:18,LiuJPCC19,quach:prr20,Manzanoarxiv20}, transport~\cite{znidaric:pre11,thingnaprb12,asadian:pre13,manzano:njp16}, and quantum information~\cite{hu:sr20,kraus:08}. The most general Markovian master equation that preserves the properties of the density matrix (positivity, Hermiticity, and trace) is the Lindblad (or Gorini-Kossakowski-Sudarshan-Lindblad, GKSL) equation~\cite{lindblad:cmp76,gorini:jmp76,breuer_02,manzano:aip20}. This equation describes the dynamics of a system under the effect of a Markovian environment. The fixed points of this dynamics have also been broadly analysed. Evans proved that~\cite{evans:jfa79} bounded systems present at least one fixed point, and that there can be more than one leading to {\it degeneracy} of the Liouvillian. 

The study of degenerated master equations has been very active during the last decade. The use of symmetries and degeneracy has been applied to reduce the dimensionality of open quantum systems~\cite{buca:njp12}, to harness quantum transport~\cite{manzano:av18}, to detect magnetic fields~\cite{thingna:njp20}, and in error correction~\cite{lieu:arxiv20}. In the timely field of quantum machine learning there are approaches to pattern retrieval by the use of degenerated open quantum systems~\cite{fiorelli:pra19}. Furthermore, the non-equilibrium properties of molecular systems have been addressed to detect symmetries and multiple fixed points~\cite{thingna:sr16}. 

In the non-degenerated case the initial condition independent steady state of a system can be obtained by numerically diagonalising the dissipative Liouvillian. Unfortunately, the degenerated case is complicated because a linear combination of fixed points is also fixed and thus there is no guarantee that the diagonalization algorithm will return the physical steady states instead of their linear combinations. Thus, the problem of degenerated Liouvillians becomes non trivial and hard to analyse numerically since the initial condition dependence cannot be easily eliminated. 

In this paper, we present a toolbox for the extraction of the physical steady-states of degenerated open quantum systems in the Lindblad form. We present three different methods, a block diagonalization, a Gramm-Schmidt-inspired orthonormalization, and a method based in large deviation theory. Each method has its own strengths and weaknesses.  To illustrate the presented methods  we apply them to a model of a ring driven out of equilibrium by two thermal baths. We analytically calculate the steady-states, for a specific choice of the parameters, by the block-diagonalization method. We discuss the phenomenology of the open quantum system as a function of its bath parameters and test the numerical methods. The minimal model allows us to analytically discuss a plethora of interesting scenarios, e.g., we find the invariant subspace of the Liouvillian can become degenerate if the bath is engineered to only pump energy into the system. In other words, even though one expects a single steady state corresponding to the invariant subspace we find multiple steady states due to the dynamical degeneration of the invariant subspace. The Gramm-Schmidt inspired method also allows us to explore the eigen-spacing statistics of the nonequilibrium steady state (NESS) and understand the signatures from the perspective of random matrix theory ~\cite{ProsenPRL13}.

The paper is organized as follows: In Sec.~\ref{sec:2} we discuss the main idea behind degenerated Liouvillians and symmetries in open quantum systems. Sec.~\ref{sec:3} is dedicated to the general formulation of the three different methods to obtain the steady states. Particularly, Sec.~\ref{sec:3A} deals with the block diagonalization approach in which the open system symmetry operators are known. In Sec.~\ref{sec:3B} we discuss the Gramm-Schmidt based orthonormalization procedure that allows us to obtain all the steady states and Sec.~\ref{sec:3C} is dedicated to the large deviation theory based method which helps obtain the non-degenerate states carrying minimum or maximum current. In Sec.~\ref{sec:4} we apply our different methods to a \emph{para}-Benzene ring, discuss analytically solvable cases, and study the eigen-spacing statistics of the NESS. Finally in Sec.~\ref{sec:5} we conclude and provide a future outlook.

\section{Degenerated Liouvillians}
\label{sec:2}

In this section we present the basics of degenerated Liouvillians and set up the notation that will be used in the paper. The main object of this study are mixed quantum states described by density matrices. If the Hilbert space of the pure states of our system is $\cH$.  A mixed state is determined by a matrix $\rho\in O(\cH)$, with $O(\cH)$ being the space of bounded operators, that fulfils two properties:
\ben
&&\text{Normalization:} \;\Tr (\rho)=1 \nonumber\\
&&\text{Positivity:}\quad \rho >0 \quad \text{i.e.,} \quad \forall \ket{\psi} \in \cH \quad \bra{\psi} \rho \ket{\psi} \geq 0.
\een
Any matrix fulfilling these two properties is considered a density matrix. Another important concept we will use is orthogonality of density matrices. Two density matrices $\rho_i$ and $\rho_j$ are considered orthogonal if  $\Tr[\rho_i \rho_j]=0$. 

In this work, we consider the dynamics of the system to be governed by the GKSL equation (see Ref.~[\onlinecite{manzano:aip20}] for an introduction), 
\ben
 \frac{d \rho(t)}{dt} &=& -i \cor{H,\rho(t)} + \sum_{i} \left( L_i\rho(t) L_i^\dagger - \frac{1}{2}\acor{\rho(t),L_i^\dagger L_i} \right), \nonumber \\
 &\equiv& \cL [\rho(t)],
 \label{eq:me}
\een
where $H$ is the Hamiltonian of our system of interest and $L_i$ are positive bound operators called ``jump operators''. Throughout this work we will set $\hbar=k_B=1$. The super-operator $\cL$ is usually named the {\it Liouville operator} of the system dynamics or just the {\it Liouvillian}. If the system pure states, $\cH$,  has a dimension $N$ the operators space dimension, $O(\cH)$ is $N^2$. As the Lindblad equation represents a map of operators, the Liouvillian $\cL$ may be represented by a matrix of dimension $N^2 \times N^2$. 

For bounded systems, Evans' theorem states that this equation has at least one fixed point~\cite{evans:jfa79}, meaning that there is at least one density matrix $\rho$ s.t. 
\be
\textrm{Re} \{\cL [\rho]\}=0. 
\ee
In most cases, there is at least one state s.t. $\cL [\rho^{{\rm SS}}]=0$. These are called steady-states and they do not evolve with time as  $d\rho^{{\rm SS}}/dt =\cL [\rho^{{\rm SS}}]=0$. Evans' theorem, as stated above, also includes the possibility of having pairs of states with zero real part but non-zero imaginary one~\cite{albert:pra14,manzano:av18}. These states are called {\it stationary coherences} and they evolve indefinitely.

The Liouvillian is a super-operator and hence to obtain its spectrum we need to map it to a matrix. The mathematical tool to do so is called the Fock-Liouville space (FLS). In the FLS, the density matrices are written as column vectors using an arbitrary map for its elements. All maps produce equivalent results and hence any choice of the map is a good choice. Once the density matrix is mapped to a column vector the Liouvillian super-operator can be written as a $N^2\times N^2$ general non-Hermitian matrix. It has both right and left eigenvectors and steady states (fixed points) correspond to the right eigenvectors with zero real eigenvalue. 

Evan's theorem also gives the conditions for obtaining a unique steady state~\cite{evans:jfa79}. This happens iff the set of operators $\{ H, L_i \}$ can generate the entire algebra of the space of bounded operators under multiplication and addition. In general, this condition is hard to prove for most systems (see Ref.~[\onlinecite{prosen:ps12}] for an example). However, when not fulfilled there are more than one steady states. This degeneracy in the Liouvillian may be related to the presence of symmetries as we discuss in the next section. 

Let's suppose that we have a degenerated Liouvillian with $M$ zero eigenvalues (we suppose there are no oscillating coherences). Each zero eigenvalue has an associated right-eigenvector that can be obtained by diagonalizing the Liouvillian expressed in the FLS. One could naively think that each of these right eigenvectors corresponds to a steady-state density matrix, but this is true only in very simple cases. In general, any linear combination of the steady-state density matrices is a right eigenvector of the Liouvillian with zero eigenvalue, but it is not necessarily a density matrix in the sense that is may not be positive. Furthermore, it is also possible that the obtained right eigenvectors do not form an orthogonal set~\footnote{Note that duality of basis ensures the left and right eigenvectors are form an orthonormal set. This does not ensure that the right eigenvectors are orthogonal amongst themselves.}, meaning that they do not belong to different invariant subspaces. Bearing these issues in mind, in the next section we propose various approaches to obtain the steady state density matrices which are independent of initial conditions in each subspace of the Liouvillian.

\section{Methods to obtain steady states}
\label{sec:3}

We present three methods to calculate the steady-states of degenerated Liouvillians, the symmetry-decomposition, the orthonormalisation and the large deviation method. Each method has its own advantages. The symmetry-based one can be applied analytically for many cases and it is numerically cheap,  but it requires full knowledge of the system's symmetries. The orthonormalisation can be applied with no previous knowledge about open system symmetries, but it's computational cost increases with the degree of degeneracy. Finally, the large deviation method does not require previous knowledge about open system symmetries and it's computationally cheap but it only gives the non-degenerated maximum and minimum current carrying states.

\subsection{Diagonalisation by symmetry-decomposition}
\label{sec:3A}
In this sub-section, we explain the relation between open system symmetry operators and multiple steady-states. We then use the knowledge of the symmetry operators and outline a procedure to obtain the steady states, some of which could have zero trace (non-physical density matrices). 

To simplify our discussion we focus on \emph{strong} open system symmetries in which there exists a unitary operator $\pi$ s.t.~\cite{buca:njp12,manzano:av18}
\be
[\pi,H]=[\pi,L_i]=0\quad \forall i.
\ee
This implies that the generators of the dissipative system dynamics $\{ H, L_i \}$ and the symmetry operator $\pi$ can be diagonalised with a common basis. Let us denote the eigenvalues of $\pi$ as $v_i=e^{i\theta_i}$, with $i\in[1,n]$ and $n$ being the number of distinct eigenvalues. Each eigenvalue can be degenerated and hence we introduce the index $d_i$ that represents the dimension of the subspace corresponding to eigenvalue $v_i$. The corresponding eigenvectors of the symmetry operator $\pi$ are $\ket{v_i^{\alpha}}$, with $i\in[1,n ]$ and $1\le \alpha\le d_i$. 

We define a super-operator $\Pi$ acting on the subspace of the bounded operators of $\cH$ as
\be
\Pi \left[x \right]\equiv \pi\cdot x \cdot\pi^{\dagger}.
\ee
The spectrum of $\Pi$ is derived from the one of $\pi$ as
\be
\Pi \left[\op{v_i^{\alpha}}{v_j^{\beta}} \right] = e^{i \left( \theta_i-\theta_j \right)} \op{v_i^{\alpha}}{v_j^{\beta}}.
\ee
Thus, the Hilbert space $\cH$ can be decomposed using the spectrum of $\pi$,
\be
\cH= \bigoplus_{i=1}^n \cH_i,
\ee
with $\cH_i=\text{span} \left\{  \ket{v_i^{\alpha}}, \alpha=1,...,d_i   \right\}$. Similarly, the space of bounded operators $\cB$ can be expanded in the eigenspace of the super-operator $\Pi$ as
\be
\cB = \bigoplus_{i,j=1}^n  \cB_{i,j},
\ee
with $ \cB_{i,j}=\text{span} \left\{\op{v_i^{\alpha}}{v_j^{\beta}},\alpha=1,\cdots,d_i;\,\beta=1,\cdots,d_j   \right\}$. Using this decomposition, it is clear that these eigenspaces are invariant under the effect of the Liouvillian $\cL  [\cB_{i,j}] \subseteq \cB_{i,j}$. This implies that the Liouvillian can be block decomposed, using the basis of $\Pi$, into $n^2$ invariant subspaces.

Normalized density matrices are only possible in the subspaces $ \cB_{i,i}$, meaning that we have at least $n$ steady states. It is also possible to find states having zero trace, belonging to the subspaces $\cB_{i,j}$ $(i\ne j)$ \cite{thingna:njp20}. These states do not represent real density matrices, but they can form linear combinations with the steady states making physical differences. Note that we use the term ``steady state'' only for the states with finite trace and corresponding to zero eigenvalue of the Liouvillian. From the above description, it is also clear that steady states corresponding to different subspaces are orthogonal to each other.

The knowledge of a strong symmetry operator $\pi$ gives us only a lower bound of the number of steady states. It is always possible that some of the blocks $\cB_{i,i}$ are further degenerated. This happens when there are $K>1$ strong symmetry operators, i.e., $\left\{ \pi^{(1)}, \dots, \pi^{(K)} \right\}$ each of them with $n^{(j)}$ ($j=1,\cdots,K$) different eigenvalues s.t.~\cite{zhang_jpa20}
\be
[\pi^{(j)},H]=[\pi^{(j)},L_i]=[\pi^{(j)},\pi^{(l)}]=0 \quad \forall (i,j,l). 
\ee
In this case we can perform the block-diagonalization of the Liouvillian using the eigenbasis of $\pi^{(1)}$, obtaining
\be
\cH=\bigoplus_{i=1}^{n^{(1)}} \cH_i.
\ee
Then each block $\cH_i$ can be further block diagonalised into a maximum of $n^{(2)}$ blocks using the eigenbasis of $\pi^{(2)}$. This can be repeated until all symmetry operators are used. Thus, since the operation of each symmetry operator not always diagonalize the Liouvillian into exactly $n^{(i)}$ blocks it is impossible to predict the total number of steady states. Thus, we can only impose bounds on the number of steady states $M$ as $\text{max} \left[n^{(i)} \right]< M< \prod_{i=1}^K n^{(i)}$.

To summarise the above outlined approach we provide an algorithm to be applied to a system having $K$ symmetry operators $\left\{ \pi^{(j)} \right\}$ ($j=1,\cdots,K$). Each of the symmetry operators $\pi^{(j)}$ have $n^{(j)}$ distinct eigenvalues with phases $\left\{ \theta^{(j)}_1,\theta^{(j)}_2, \dots, \theta^{(j)}_{n^{(j)}}\right\}$. As the symmetry operators commute with each other we can define a common eigenbasis of all of them. The eigenbasis can be defined by the eigenvectors $\left\{  \ket{v_{\theta^{(1)}_{i_1}, \theta^{(2)}_{i_2},\cdots,\theta^{(K)}_{i_K}  }^{\alpha}} \right\}$, where $i_j \in [1,n^{(j)}]$, and $\alpha$ stands for the degeneracy of the subspace determined by the eigenvalues $\vec{\theta}_{\vec{i}}=\left\{  \theta^{(1)}_{i_1}, \theta^{(2)}_{i_2},\cdots,\theta^{(K)}_{i_K}   \right\}$ where $\vec{i}=\{i_1,i_2,\cdots,i_K\}$ and each element $i_j$ of $\vec{i}$ is associated with the same element $\theta^{(j)}$ of $\vec{\theta}$. This means that each vector $\ket{v_{\vec{\theta}_{\vec{i}}}^{\alpha}}$ is an eigenvector of each symmetry operator $\pi^{(j)}$, i.e.,
\ben
\pi^{(j)}\ket{v_{\vec{\theta}_{\vec{i}}}^{\alpha}}  = \theta_{i_j}^{(j)} \ket{v_{\vec{\theta}_{\vec{i}}}^{\alpha}}.
\een
The eigenbasis of the corresponding super-operators $\Pi^{(j)}$ is naturally given by the elements $\left\{  \ket{v_{\vec{\theta}_{\vec{i}}}^{\alpha}}  \bra{v_{\vec{\theta}_{\vec{i^{\prime}}}}^{\beta}} \right\}$. The method to obtain the steady sates of the degenerated Liouvillian, if we know its symmetry operators, is then: 
\begin{enumerate}
\item Find the common eigenbasis of all the symmetry operators $\left\{ \pi^{(j)}\right\}$.  
\item Calculate the eigenvalues of the symmetry operators corresponding to the elements of the basis, obtaining a classification of the form $\ket{v_{\vec{\theta}_{\vec{i}}}^{\alpha}} $.
\item Order the elements of the basis by grouping all the vectors with the same eigenvalues. 
\item Change the Liouvillian to the new basis. A block-diagonal structure arises. 
\item Diagonalise each block of the new basis. Any eigenvector with a zero eigenvalue corresponds to a steady state. Note that the dimension of the blocks are smaller than the dimension of the Liouvillian and, therefore, the eigenvectors of the blocks do not represent density matrices by themselves.
\item Increase the dimension of the eigenvectors of each block by adding $0$'s to complete the dimension. 
\item Change back to the original basis. 
\end{enumerate} 

\subsection{Diagonalisation by orthonormalization}
\label{sec:3B}
In the last sub-section we dealt with the ideal scenario in which all the strong symmetry operators were known. In complex many-body open quantum systems knowing all the strong symmetry operators is highly non-trivial and the problem can become even more complicated if \emph{weak} symmetry~\cite{buca:njp12} is degenerating the Liouvillian. In this case, our starting point could be a set of $M$ linearly independent right eigenvectors of the Liouvillian which correspond to zero eigenvalue. One could naively expect that these operators are indeed the density matrices corresponding to the fixed points of the Liouvillian, but this is not the general case. In most cases, the diagonalization algorithm will give us a set of operators that are neither positive nor orthogonal to each other. Thus, in this sub-section we explain our second method to reconstruct the density matrices from such a set. This method was first presented in Ref.~[\onlinecite{thingna:njp20}] and it does not require any pre-requisite knowledge of the strong or weak symmetry operators.

Having this objective in mind the question we ask is: If we have a set of $M$ zero eigenvalue eigenvectors of $\cL$ that are linearly independent $\left\{ \tilde{\rho}_i \right\}$, how can we reconstruct $M$ positive density matrices $\left\{ \rho_i\right\}$ with the following properties: 
\ben
\cL[\rho_i] &=& 0 \quad\forall i,  \\
%\text{Tr}(\rho_i) &=& 1\quad\forall i \\
\text{Tr}[\rho_i \rho_j] &=& 0\quad\forall i\ne j.
\een

We will address this problem by a two-step approach. First, we construct a set of orthogonal matrices. To construct the orthonormal set we start by applying an orthogonalisation process based on Gramm-Schmidt algorithm. To begin, we form a set of Hermitian matrices $\left\{ \rho^H_i\right\}$ from the original set,
\be
\rho^H_i =  \tilde{\rho}_i + \tilde{\rho}_i^\dagger. 
\label{eq:herm}
\ee
Then we use these Hermitian matrices $\left\{ \rho^H_i\right\}$ to construct a set of orthogonal Hermitian matrices by applying 
\begin{eqnarray}
 \rho_1^O &=& \rho_1^H, \nonumber\\
 \rho_2^O &=& \rho_2^H- \frac{\Tr[\rho_1^O \; \rho_2^H ]}{\Tr[\rho_1^O \;\rho_1^O ]} \rho_1^O, \nonumber\\
 \rho_3^O &=& \rho_3^H- \frac{\Tr[\rho_1^O \; \rho_3^H]}{\Tr[\rho_1^O \;\rho_1^O ]} \rho_1^O - \frac{\Tr[\rho_2^O \; \rho_3^H]}{\Tr[\rho_2^O \;\rho_2^O]} \rho_2^O, \nonumber\\
&  \vdots&  \nonumber\\
 \rho_M^O &=& \rho_M^H - \sum_{j=1}^{M-1}  \frac{\Tr [\rho_j^O \; \rho_N^H]}{\Tr[\rho_j^O \;\rho_j^O]} \rho_j^O. 
\end{eqnarray}
The orthonormalization process preserves Hermiticity and it trivially follows that the set $\left\{ \rho^O_i \right\}$ fulfil the orthogonality relation 
\be
\Tr [ \rho_i^O  \rho_j^O ] =  0 \quad \text{if} \quad i\neq j.
\ee
This is a set of eigenmatrices of the Liouvillian with zero eigenvalue in which every matrix is Hermitian and orthogonal to each other. The only remaining issue is that these matrices may not be semi-positive definite, meaning that they may have negative eigenvalues. To address this issue, we first define the {\it positivity} functional, $P$, of a set of $M$ Hermitian operators, $\left\{ A_i\right\}_{i=1}^M $, of dimension $N$ (same as the dimension of density matrices) as 
\be
P\left[ \left\{ A_i \right\} \right]= \sum_{i=1}^M \sum_{j=1}^N v_{j}^{A_i} - \left| v_{j}^{A_i} \right|,
\ee
with $v_{j}^{A_i}$ being the $j$th eigenvalue of operator $A_i$. It is clear that this measure is equal to zero iff all the matrices of the set $\left\{ A_i\right\}_{i=1}^M$ are semi-positive definite. As the set of matrices $\left\{ \rho^O_i \right\}_{i=1}^{M}$ are orthogonal and a linear combination of positive matrices, we may find a unitary operator, $U$, that transforms this set to a zero eigenvalue positive orthogonal matrices $\left\{ \rho_i^P \right\}$. To do so, we first write the original set as a column vector 
\be
|\rho^O\rangle\rangle \equiv
\left(
\begin{array}{c}
\rho_{1}^O \\
\rho_{2}^O\\
\vdots \\
\rho_{M}^O
\end{array}
\right). 
\ee
As we want to preserve orthogonality, we need to apply a unitary operator to the vector $|\rho^O\rangle\rangle$. This transformation can be described by a set of $(M^2-M)/2$ Euler angles, $\vec{\chi}=\left\{ \chi_1, \, \chi_2 ,\dots, \, \chi_{\frac{M^2-M}{2}}  \right\}$. For a specific choice of the Euler angles we can define the new vector of matrices $|\rho(\vec{\chi})\rangle\rangle = U(\vec{\chi}) |\rho^O\rangle\rangle$, corresponding to the set of matrices $\left\{ \rho_i(\vec{\chi}) \right\}$. 

In order to find the correct choice of the angles that performs the correct transformation we need to maximise the functional 
\be
F\left[  \left\{ \rho_i(\vec{\chi}) \right\} \right] = \sum_{i=1}^M \sum_{j=1}^N v_{j}^{\rho_i(\vec{\chi})}  - \left| v_{j}^{\rho_i(\vec{\chi})}  \right|,
\label{eq:functional}
\ee
with respect to the various Euler angles. Thus, we can obtain a set of orthogonal semi-definite positive zero eigenvalue right-eigenvector matrices $\left\{ \rho_i^P \right\}$. These obtained matrices need not be normalized and this can be easily achieved by transforming $\rho_i=\rho_i^P/\Tr[\rho_i^P]$ for all the matrices that have $\Tr[\rho_i^P] \neq 0$.

The above described method can be summarised as follows:
\begin{enumerate}
\item Obtain a set of Hermitian matrices by applying Eq.~\eqref{eq:herm} and obtaining the set  $\left\{ \rho^H_i\right\}$ .
\item Construct a set of orthogonal matrices,  $\left\{ \rho^O_i \right\}$, by applying a Gram-Schmidt method for density matrices. 
\item Find the rotation angles, $\vec{\chi}=\left\{ \chi_1, \, \chi_2 ,\dots, \, \chi_{\frac{M^2-M}{2}}  \right\}$, by maximising the functional, Eq.~\eqref{eq:functional}. 
\item Apply the rotation $U(\vec{\chi})$ to obtain the orthonormal semi-positive definite Hermitian set of matrices $\left\{ \rho_i^P \right\}$.
\item Renormalise by doing $\rho_i=\rho_i^P/\Tr[\rho_i^P]$ for all the matrices that have $\Tr[\rho_i^P] \neq 0$.
\end{enumerate}

\subsection{Diagonalisation by large deviations}
\label{sec:3C}
In this sub-section we describe a method to obtain some of the steady states by a single diagonalization of the Liouvillian, making it much simpler than the previous methods. On the other hand, it can be applied only in some cases and it allows us to obtain only some of the states. The method is based on the study of the thermodynamic currents and it was first presented in Ref.~[\onlinecite{manzano:prb14}] (see Ref.~[\onlinecite{manzano:av18}] for a more detailed discussion). Here we focus only on the description of this approach and its applicability. 

We consider a system connected to several incoherent channels that allow the exchange of quanta between the system and an environment. This allows us to divide the super-operator $\cL$ from Eq.~\eqref{eq:me} into three parts 
\be
\cL=\cL_{-1} + \cL_{0}+ \cL_{+1},
\ee
where the subscripts indicate the number of excitations introduced/removed from the system by the environment. Of course, there could be more exotic environments that exchange more than one excitation but for the sake of simplicity we will not consider this possibility. Next, we define the system density matrix conditioned on a fixed number of excitations $Q$ as $\rho_Q(t) \equiv \Tr_Q[\rho(t)]$ where $\Tr_Q$ is partial trace over the manifold containing $Q$ excitations. Thus, the evolution of $\rho_Q(t)$ is governed by
\be
\frac{d \rho_Q(t)}{dt} = \cL_{-1}  [\rho_{Q+1}(t)] + \cL_{0}  [\rho_Q(t)] + \cL_{+1}  [\rho_{Q-1}(t)].
\label{eq:meQ}
\ee
This gives a hierarchy of equations that can be unravelled using the Laplace transform 
\be
\rho_\lambda (t) = \sum_{Q=-\infty}^\infty \rho_Q (t) e^{-\lambda Q}, 
\ee
which when applied to Eq.~\eqref{eq:meQ} gives a set of independent equations
\begin{eqnarray}
\frac{d \rho_\lambda(t)}{dt} &=&  e^{\lambda} \cL_{-1}  [\rho_{\lambda} (t)] + \cL_{0}  [\rho_{\lambda} (t)] + e^{-\lambda} \cL_{+1}  [\rho_{\lambda}(t)] \nonumber \\
& \equiv &  \cL_{\lambda} [\rho_{\lambda}(t)].
\end{eqnarray}
where $\lambda$ in known as the {\it counting field}. For the Lindblad equation that takes the form of Eq.~\eqref{eq:me}, we have the correspondence
\ben
\cL_{-1}[\rho(t)] &=& L_i \rho(t) L_i^\dagger \nonumber\\
\cL_{+1}[\rho(t)] &=& L_j \rho(t) L_j^\dagger \nonumber\\
\cL_{0} [\rho(t)] &=& -i \left[  H,\rho(t) \right]  \\
&&+ \sum_{k \neq i,j} L_k \rho(t) L_k^\dagger - \frac{1}{2} \sum_k \left\{ L_k L_k^\dagger,\rho(t) \right\}, \nonumber
\een
where the index $i/j$ stand for the incoherent channels that extract/inject excitations in the system. The probability of finding the system in a state with $Q$ excitations is $P_Q(t)=\Tr[\rho_Q(t)]$, and 
\be
Z_{\lambda}(t) \equiv \Tr[\rho_\lambda(t)] = \sum_{Q=-\infty}^\infty P_Q(t) e^{-\lambda Q},
\ee
is known as the generating function of the current probability distribution. This generating function follows a {\it large deviation principle} and for long times it scales as 
\be
Z_{\lambda}(t) \sim e^{t \mu(\lambda)},
\ee
where $\mu(\lambda)$ is called the current {\it Large Deviation Function} (LDF). It can be calculated as the highest eigenvalue of the tilted super-operator $\cL_{\lambda}$. As $Z_{\lambda}(t)$ is the current moment generating function, the LDF $\mu_{\lambda}$ corresponds to the cumulant generating function of the current distribution. Therefore, the average current can be calculated as  
\be
\langle  \dot{Q} \rangle = \lim_{t\to \infty} \left.  \frac{1}{t} \frac{\partial Z_\lambda (t)}{\partial \lambda} \right|_{\lambda=0} = \left. \frac{\partial \mu(\lambda)}{\partial \lambda}\right|_{\lambda=0}. 
\ee
If $\left| \lambda \right|<<1$ we can expand the LDF as 
\be
\left. \mu(\lambda)\right|_{\lambda\to 0} \sim \mu(0) + \left. \frac{\partial \mu(\lambda)}{\partial \lambda}\right|_{\lambda=0} = \langle \dot{Q} \rangle. 
\ee
Therefore, if the Liouvillian is degenerated and the different steady states have different average currents the LDF $\mu(\lambda)$ will have a non-analytic behaviour around $\lambda=0$ in the form 
\be
\mu(\lambda)=
\left\{
\begin{array}{cc}
+ \left| \lambda \right|  \langle\dot{Q} \rangle_\text{max} & \text{for }\lambda \to 0^- \\
- \left| \lambda \right|  \langle \dot{Q} \rangle_\text{min} & \text{for }\lambda \to 0^+
\end{array}
\right.
\ee
This allows us to calculate the steady-states corresponding to the maximum and minimum currents as long as they are not degenerated. The method may be summarised as follow: 
\begin{enumerate}
\item Calculate the highest eigenvalue $\mu(\lambda) $(and its corresponding eigenvector $\rho_\lambda$) of the modified Liouvillian $\cL_\lambda$.
\item Take the limits $\rho'_\text{min}= \lim_{\lambda\to 0^+} \rho_\lambda$ and $\rho'_\text{max}= \lim_{\lambda\to 0^-} \rho_\lambda$.
\item Renormalize, obtaining $\rho_\text{min} = \rho'_\text{min}/\Tr[\rho'_\text{min}]$ and  $\rho_\text{max} = \rho'_\text{max}/\Tr[\rho'_\text{max}]$.
\end{enumerate}

To summarise this section, we have introduced three different methods using which we can obtain the steady states for an open quantum system with a degenerated Liouvillian. The first method described in Sec.~\ref{sec:3A} is the most general approach, but requires the knowledge of symmetry operators which are usually difficult to obtain. The second approach (Sec.~\ref{sec:3B}) could be easily implemented computationally and does not require any knowledge of the symmetry operators. Although this seems most beneficial, with increase in the degree of degeneracy the computational cost increases substantially due to the minimization procedure to find the optimal Euler angles. The final method is the easiest computationally (Sec.~\ref{sec:3C}), but is limited to class of nonequilibrium systems and can be used to obtain only a subset of the steady states.

\section{Example: \emph{para}-Benzene ring}
\label{sec:4}
\begin{figure}
\begin{center}
\includegraphics[width=\columnwidth]{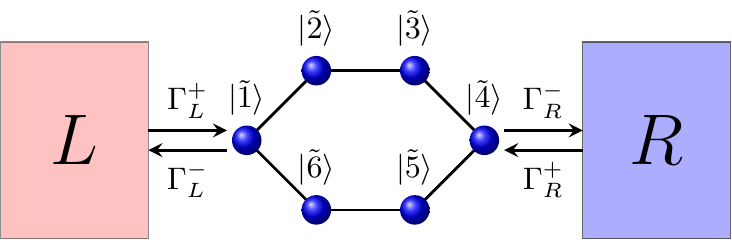}
\end{center}
\caption{Illustration of the para-Benzene-type system with 6 sites connected to two incoherent baths (red and blue rectangles) at different temperatures $T_L$ and $T_R$. The para-Benzene system exchanges energy with the left $L$ and right $R$ baths due to the pumping rates $\Gamma^+$ and dumping rates $\Gamma^-$. The tilde basis is the original site representation.}
\label{fig:ring}
\end{figure}

The methods presented can deal with a wide variety of scenarios and in order to illustrate these we use the example of a \emph{para}-Benzene ring connected to two reservoirs as illustrated in Fig.~\ref{fig:ring}. We restrict to the single-excitation picture and consider the Hilbert space to be spanned by the site basis $\left\{\ket{\tilde{i}} \right\}_{i=1}^6$ plus a ground state $\ket{\tilde{0}}$ to allow interactions with the reservoir. The system Hamiltonian takes the form
\be
H = J \sum_{\tilde{n}=1}^6 \op{\tilde{n}}{\widetilde{n+1}} + {\rm H.c.}.
\ee
with $|\tilde{7}\rangle=|\tilde{1}\rangle$. The system is boundary driven by two incoherent baths connected to sites $1$ and $4$. The baths exchange energy and excitations with the system via the jump operators 
\ben
\label{eq:jump}
L_1= \sqrt{\Gamma_L^+} \op{\tilde{1}}{\tilde{0}}, \quad
L_2= \sqrt{\Gamma_L^-} \op{\tilde{0}}{\tilde{1}}, \nonumber\\
L_3= \sqrt{\Gamma_R^+} \op{\tilde{4}}{\tilde{0}}, \quad 
L_4= \sqrt{\Gamma_R^-} \op{\tilde{0}}{\tilde{4}},
\een
where $\Gamma_{\rm x}^{+(-)}\geq 0$ are the pumping (dumping) rates for the ${\rm x}$th bath (${\rm x} = L$ or $R$). All properties of the baths are encoded in these rates and we will not consider any specific form herein. For this simple ring structure, there is only one open system symmetry operator given by,
\be
\pi = \sum_{i=0,1,4}\op{\tilde{i}}{\tilde{i}} +\op{\tilde{2}}{\tilde{6}} + \op{\tilde{6}}{\tilde{2}} + \op{\tilde{3}}{\tilde{5}} + \op{\tilde{5}}{\tilde{3}}.
\ee
The unitary operator $\pi$ has two eigenvalues $+1$ and $-1$ and the transformation matrix ${\rm T}$ to change basis from the site representation to the eigenvectors $|i\rangle$ of $\pi$ reads,
\ben
\label{eq:transformation}
{\rm T}\,|\tilde{i}\rangle &=&|i\rangle \nonumber \\
{\rm T} &=& \sum_{i=0,1,4}\op{\tilde{i}}{\tilde{i}} +\frac{1}{\sqrt{2}}\sum_{i=2,3}\op{\tilde{i}}{\tilde{i}} -\frac{1}{\sqrt{2}}\sum_{i=5,6}\op{\tilde{i}}{\tilde{i}} \nonumber \\
&&+\frac{1}{\sqrt{2}}\left(|\tilde{2}\rangle\langle\tilde{6}| + |\tilde{3}\rangle\langle\tilde{5}| + \mathrm{H.c.}\right).
\een
The ground $|0\rangle$ and symmetric states $|i\rangle$ ($i=1,\cdots,4$) have eigenvalue $+1$ whereas the anti-symmetric states $|i\rangle$ ($i=5,6$) correspond to eigenvalue $-1$. The transformation matrix does not affect the ground ($\tilde{0}$) and \emph{edge} sites ($\tilde{1}$ and $\tilde{4}$) which are connected to the baths but only transforms the \emph{bulk} sites ($\tilde{2}$, $\tilde{3}$, $\tilde{5}$, and $\tilde{6}$).

The system Hamiltonian in the transformed basis takes the form
\begin{equation}
\label{eq:Htrans}
H = \sqrt{2} J \left( \op{1}{2} + \op{3}{4} \right)+ J \left(  \op{2}{3} + \op{5}{6} \right) +\mathrm{h.c.},
\end{equation}
which is block diagonal since the ground and symmetric subspace ($|0\rangle, \cdots, |4\rangle$) does not interact with the anti-symmetric one ($|5\rangle$ and $|6\rangle$). Since the transformation does not affect the ground state and the edge sites, there is no entanglement generated in the jump operators and they remain the same form as Eq.~\eqref{eq:jump} with $|\tilde{i}\rangle \rightarrow |i\rangle$.

Given the block diagonal form of the system Hamiltonian and the jump operators confined to the ground and symmetric subspace we can split the system space into the subspace of the ground state ($\cH_g$ with 1 state), symmetric states ($\cH_s$ with 4 states), and anti-symmetric states ($\cH_a$ with 2 states). Thus, the system Hamiltonian can be decomposed into a $3\times 3$ matrix that takes the form
\be
H= \left(
\begin{array}{ccc}
0 & 0 & 0\\
0 & H_{ss} & 0\\
0 & 0 & H_{aa}\\
\end{array}
\right).
\ee
In this representation the sum of the jump operators takes the form
\ben
\sum_{i=1}^{4}L_i= \left(
\begin{array}{ccc}
0 & L_- & 0\\
L_+ & 0 & 0\\
0 & 0 & 0\\
\end{array}
\right)
\een
with $L_+=L_1 + L_3$ representing the net pumping operator and $L_-=L_2+L_4$ being the net dumping operator. The Lindblad equation~\eqref{eq:me} then separates out for each sub block and the resultant equations read
\ben
\label{eq:symmsub}
\frac{d\rho_{gg}(t)}{dt} &=& -\frac{1}{2} \{N_+, \rho_{gg}(t) \}+L_- \rho_{ss}(t) L_-^\dagger, \nonumber \\
\frac{d\rho_{ss}(t)}{dt} &=& - i [ H_{ss},\rho_{ss}(t)] -\frac{1}{2} \{ N_-, \rho_{ss}(t)  \} \nonumber \\
&&+ L_+\rho_{gg}(t) L_+^\dagger, \\
\frac{d\rho_{gs}(t)}{dt}&=& i\rho_{gs}(t) H_{ss} -\frac{1}{2} \rho_{gs}(t) N_- - \frac{1}{2} N_+ \rho_{gs}(t), \nonumber \\
\frac{d\rho_{ga}(t)}{dt} &=& i\rho_{ga}(t) H_{aa}  - \frac{1}{2} N_+ \rho_{ga}(t),  \nonumber \\
\label{eq:decay}
\frac{d\rho_{sa}(t)}{dt} &=& - i \left(  H_{ss} \rho_{sa}(t) - \rho_{sa}(t) H_{aa}  \right) - \frac{1}{2} N_- \rho_{sa}(t), \\
\label{eq:asymmsub}
\frac{d\rho_{aa}(t)}{dt} &=& -i [ H_{aa}, \rho_{aa}(t) ],
\een
with $N_+ = L_+^{\dagger}L_+$ and $N_- = L_-^{\dagger}L_-$ being positive operators and $\rho_{{\rm x},{\rm y}}(t) = \rho_{{\rm y},{\rm x}}^{\dagger}(t)$ ($\{{\rm x},{\rm y}\} = g,s,a$). The cross-subspaces, i.e., $\rho_{{\rm x},{\rm y}}(t)~\forall{\rm x} \neq {\rm y}$, the reduced density matrix $\rho_{{\rm x},{\rm y}}(t)$ decays exponentially as can be seen from Eq.~\eqref{eq:decay}. 

Thus in the steady state only the the diagonal components of the reduced density matrix survive and we now focus on the anti-symmetric subspace whose evolution is described by Eq.~\eqref{eq:asymmsub}. Clearly, this describes coherent evolution and thus the anti-symmetric subspace is a decoherence free subspace. The eigenvectors of $H_{aa}$ ($2\times 2$ matrix) can be easily obtained and are given by,
\ben
\ket{\psi_1} &=& \frac{1}{\sqrt{2}} \left(  \ket{5} + \ket{6}  \right),\nonumber \\
\ket{\psi_2} &=& \frac{1}{\sqrt{2}} \left(  \ket{5} - \ket{6}  \right).
\een
If we initiate our system in any one of these states it will not evolve in time and hence from the perspective of the general Lindblad equation both of these pure states are steady states. In other words, the dark states 
\be
\label{eq:darkstates}
\rho^{{\rm DS}}_1 = \op{\psi_1}{\psi_1}\quad{\rm and}\quad\rho^{{\rm DS}}_2 = \op{\psi_2}{\psi_2}
\ee
are zero current carrying steady states. The cross combination of these states display oscillating behaviour and are known as oscillating coherences~\cite{albert:pra14} whose state has zero trace $\rho^{{\rm OC}}(t)=e^{-i2Jt}\op{\psi_1}{\psi_2}+e^{i2Jt}\op{\psi_2}{\psi_1}$. The frequency of the oscillations is given by the difference of the eigenvalues of $H_{aa}$. Thus, the existence of decoherence free subspaces always gives us $L$ number of steady states, where $L$ is the dimension of the decoherence free subspace, and $L$ pairs of eigenvalues of the Liouvillian with zero real part but finite imaginary part known as oscillating coherences.

The reduced density matrix for the subspaces belonging to the ground and symmetric states obeys coupled first order differential equations [see Eq.~\eqref{eq:symmsub}] which is impossible to solve analytically. In general, this set up has \emph{three} steady states; one from the ground and symmetric subspace and two from the anti-symmetric subspace described above. In specific scenarios, wherein the effect of the bath can be simplified we can obtain analytic solutions as described below.

\subsection*{Equilibrium}
We can simplify our problem by considering that the pumping (dumping) rates of both baths are the same, i.e., $\Gamma^{+}_L = \Gamma^{+}_R = \Gamma$ and $\Gamma^{-}_L = \Gamma^{-}_R=\gamma$. In this case, the equilibrium steady state is given by,
\be
\rho^{{\rm EQ}}_3 = \frac{\gamma}{\gamma + 4\Gamma}\ket{0}\bra{0} +\frac{\Gamma}{\gamma + 4\Gamma}\sum_{i=1}^{4}\ket{i}\bra{i}.
\ee
If the baths were ideal sinks $\Gamma = 0$ (zero temperature baths) or pumping and dumping at the same rate $\gamma=\Gamma$ (infinite temperature baths)  we obtain the physically intuitive results of either being localized in the ground state or all states being equally populated. Note here that in the general equilibrium scenario we do not obtain the canonical Gibbs state because the jump operators in our Lindblad equation are resonantly being coupled to the ground state $\tilde{0}$ and either site $\tilde{1}$ or $\tilde{4}$. Such a resonant coupling does not allow the dissipator to mix \emph{all} the energy levels which is a crucial requirement to obtain a Gibbsian at equilibrium.

\begin{figure}
\begin{center}
\includegraphics[width=\columnwidth]{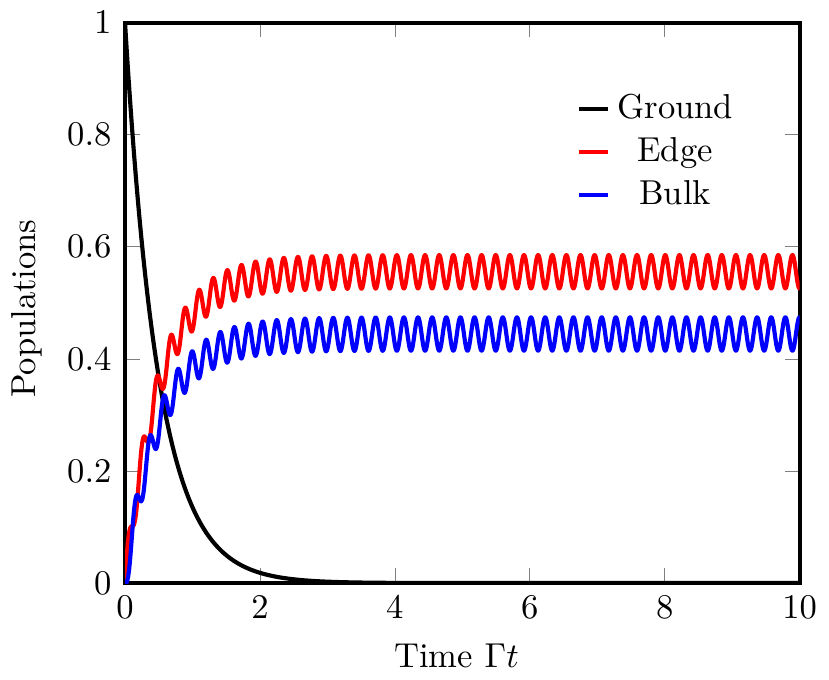}
\end{center}
\caption{Populations as a function of time $t$ for the case of pure pumping $L^-_i=0$. The system exhibits a dynamical decoherence free subspace due to which we obtain multiple steady states even in the absence of strong or weak open system symmetries. The symmetric subspace is invariant and only in the limit $t\rightarrow \infty$ the invariant symmetric subspace becomes decoherence free. The ground state population is $\bra{\tilde{0}}\rho(t)\ket{\tilde{0}}$, edge state population is $\rho_{\text{edge}}(t) = \sum_{i=1,4}\bra{\tilde{i}}\rho(t)\ket{\tilde{i}}$, and bulk state population is $\rho_{\text{bulk}}(t) = \sum_{i=2,3,5,6}\bra{\tilde{i}}\rho(t)\ket{\tilde{i}}$. All individual sites in the bulk or edge have the same populations due to the open system symmetries and the difference in the bulk and edge site populations is due to the symmetries in $H_{ss}$. The pumping rate for both baths $\Gamma^+_{\rm x}=\Gamma= 0.1$ and the hopping $J=1$.}
\label{fig:DynSymm}
\end{figure}
\subsection*{Ideal source}
In another extreme scenario when the baths are an ideal source such that $\Gamma^{+}_L = \Gamma^{+}_R = \Gamma$ and $\Gamma^{-}_L = \Gamma^{-}_R=0$ the dynamical equations of ground symmetric subspace [Eq.~\eqref{eq:symmsub}] simplify as,
\ben
\label{eq:idsource1}
\frac{d\rho_{gg}(t)}{dt} &=& -2\Gamma \rho_{gg}(t), \\
\label{eq:idsource2}
\frac{d\rho_{ss}(t)}{dt} &=& -i[H_{ss},\rho_{ss}(t)] + \Gamma\rho_{gg}(t)\sum_{i,j=1,4}\ket{i}\bra{j}.
\een
The equation for $\rho_{gg}(t)$ can be solved analytically giving an exponentially decaying solution $\rho_{gg}(t) = \exp[-2\Gamma t]\rho_{gg}(0)$ with $\rho_{gg}(0)$ being the initial condition. In the long-time limit $\rho_{gg}=0$, which is expected since the baths only pump excitations from the ground state to the ring. In this long-time limit, it is clear from Eq.~\eqref{eq:idsource2} that $\rho_{ss}(t)$ obeys an oscillatory coherent evolution. Thus, in this ideal source limit, we obtain more than three steady states (six in particular): the anti-symmetric subspace is not affected by this analysis and hence gives the two steady states as explained above, whereas the ground and symmetric subspace now give \emph{four} (dimension of $H_{ss}$) steady states using the same arguments we provided for the coherent evolution in the anti-symmetric subspace analysis. Note here that the emergence of these extra steady states is not due to the open system symmetries but because there was a dynamical restoration of Hamiltonian symmetries in the long-time limit. Thus, in general the existence of multiple steady states need not be rooted in open system symmetries (as usually believed), but could arise due to the peculiar properties of the baths.

We illustrate this evolution for the real-space populations in Fig.~\ref{fig:DynSymm}. The ground state (black solid line) population decays exponentially as expected and the populations of the edge ($\rho_{\text{edge}}(t) = \sum_{i=1,4}\bra{\tilde{i}}\rho(t)\ket{\tilde{i}}$, red solid line) and bulk ($\rho_{\text{bulk}}(t) = \sum_{i=2,3,5,6}\bra{\tilde{i}}\rho(t)\ket{\tilde{i}}$, blue solid line) sites oscillate indefinitely. The oscillations of the edge and bulk are out of phase and the difference in their amplitudes is due to the symmetries in $H_{ss}$, which has different weights for the connections between the edges and the bulk sites [see Eq.~\eqref{eq:Htrans}].

\subsection*{Ideal sink and source}
Next we turn our attention to systems in nonequilibrium. The simplest case which yields analytic results is when one of the baths is an ideal sink $\Gamma_L^+=0$, $\Gamma_L^-=\gamma$ whereas the other is an ideal source $\Gamma_R^+=\Gamma$, $\Gamma_R^-=0$. Unlike the ideal source scenario, in which the ground state gets depleted leading to dynamical restoration of Hamiltonian symmetries, in this case the ideal sink would re-populate the ground state ensuring a current carrying NESS exists. The ground and symmetric subspace have only one NESS given by,
\begin{widetext}
\ben
\rho^{{\rm NESS}}_3 &=& \frac{1}{1+\frac{4\Gamma}{\gamma}+\frac{9\gamma\Gamma}{16J^2}}\left\{\ket{0}\bra{0} + \frac{\Gamma}{\gamma}\ket{1}\bra{1} + \left(\frac{\Gamma}{\gamma}+\frac{\gamma\Gamma}{8J^2}\right)\ket{2}\bra{2} + \left(\frac{\Gamma}{\gamma}+\frac{\gamma\Gamma}{4J^2}\right)\ket{3}\bra{3} + \left(\frac{\Gamma}{\gamma}+\frac{3\gamma\Gamma}{16J^2}\right)\ket{4}\bra{4}\right.\nonumber \\
&&\left.-i\frac{\Gamma}{2\sqrt{2}J}\left(\ket{1}\bra{2}+\sqrt{2}\,\ket{1}\bra{4}+\frac{1}{\sqrt{2}}\ket{2}\bra{3}-i\frac{4J}{\gamma}\ket{2}\bra{4}+\ket{3}\bra{4}+{\rm H.c.}\right)\right\}.
\een
\end{widetext}
The steady-state excitonic currents for the ideal sink source scenario 
\ben
I_{L}&=&\Tr [L_1^{\dagger}L_1\rho^{{\rm NESS}}_3]-\Tr [L_2^{\dagger}L_2\rho_3^{{\rm NESS}}] \nonumber \\
&=& \frac{\Gamma}{1+\frac{4\Gamma}{\gamma}+\frac{9\gamma\Gamma}{16J^2}}
\een

\begin{figure}[b!]
\begin{center}
\includegraphics[width=\columnwidth]{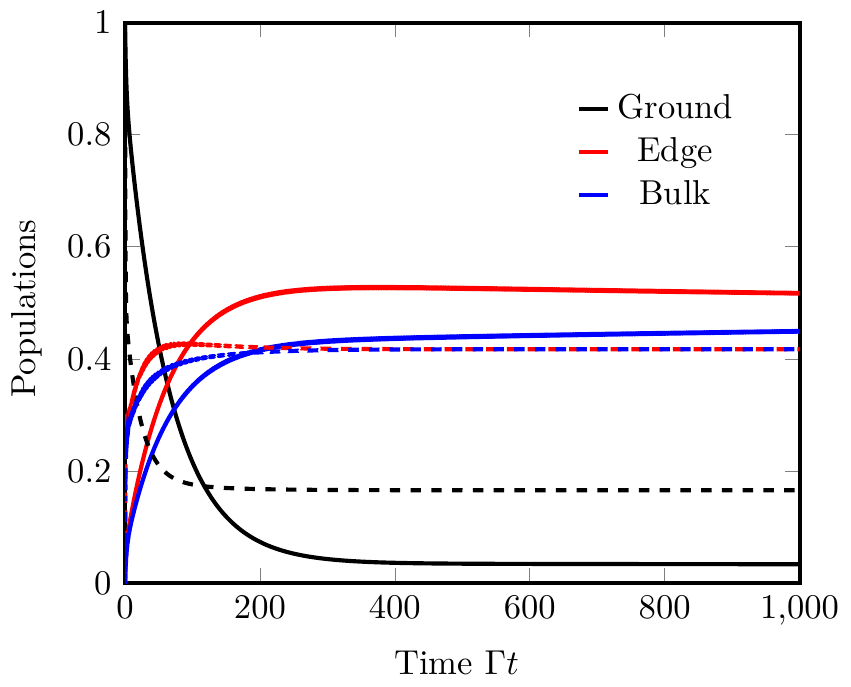}
\end{center}
\caption{Populations as a function of time $t$ for the general nonequilibrium scenario. Solid lines are for the case of low temperature with $T_L=0.25$ and $T_R=0.5$, whereas dashed lines are for the high temperature regime with $T_L=1$ and $T_R=2$. The individual edge and bulk sites (same as that defined in the caption of Fig.~\ref{fig:DynSymm}) have the same populations due to open system symmetries. At low temperatures, the edge and bulk populations are distinct exhibiting the same symmetry governed by $H_{ss}$ (same as Fig.~\ref{fig:DynSymm}). At high temperatures, the Hamiltonian symmetry is broken and the bulk and edge site populations become equal after a short transient. The hopping is chosen to be $J=1$ and the rates obey local-detailed balance, $\Gamma_{{\rm x}}^{+} = \Gamma\omega_0 n(T_{{\rm x}},\omega_0)/2$ and $\Gamma_{{\rm x}}^{-} = \Gamma\omega_0 [1+n(T_{{\rm x}},\omega_0)]/2$ with ${\rm x} = L,R$, $n(T,\omega_0) = [\exp[\omega_0/T]-1]^{-1}$ being the Bose-Einstein distribution, $\omega_0 =1$ being the system-bath resonant frequency, and $\Gamma=0.1$ the system-bath coupling strength.}
\label{fig:HighLowTemp}
\end{figure}
\subsection*{General case}
In the general nonequilibrium case it is not possible to solve the differential equations exactly and hence we solve these numerically and display the dynamics in Fig.~\ref{fig:HighLowTemp}. The solid lines in Fig.~\ref{fig:HighLowTemp} are for the low temperature regime in which we find that the edge (red lines) and bulk (blue lines) state populations are different. The difference in the populations can be attributed to the symmetries of the symmetric subspace Hamiltonian $H_{ss}$ (recall a similar behaviour was observed in the ideal-source case). At low temperatures, the bath should not affect the system dramatically and thus the Hamiltonian symmetries should be respected. On the other hand, at high temperatures [Fig.~\ref{fig:HighLowTemp} dashed lines] the dissipative baths completely alter the system dynamics and hence in this case we do not see any signatures of the $H_{ss}$ symmetries being preserved. In fact, at high temperatures the edge and bulk populations become equal after a short transient indicating a equal distribution of the excitation among the edge and bulk.
	
\subsection*{Eigenspacing statistics of NESS}
There are several scenarios in which knowing the NESS for a degenerated Liouvillian could be useful. In this subsection we focus on the timely example of studying the eigenspacing statistics of the NESS as first proposed in Ref.~[\onlinecite{ProsenPRL13}]. Recently, there has been a surge in understanding the universal properties of a dissipative open quantum system mostly restricted to the spectra of a non-degenerated Liouvillian~\cite{DenisovPRL19,DavidePRL20,ProsenPRX20}. The idea is to observe universal features based on statistical correlations between the eigenvalues of the Liouvillian or the NESS. For closed Hamiltonian systems, there is a deep connection between the quantum chaos conjecture~\cite{Berry77, BerryAP81} and the statistical correlations of the eigenvalues which is described by random matrix theory \cite{Mehta}. However, for open quantum systems very little is known in this direction.

Unlike closed systems, for a complex many-body open quantum system evaluating the entire spectra of the Liouvillian can be computationally expensive since its corresponding matrix dimension scales as $N^2 \times N^2$ (recall that $N$ is the dimension of the system Hilbert space). For degenerated Liouvillians, most studies are restricted up to $N\approx 250$. On the other hand, since the NESS is the eigenvector corresponding to the zero eigenvalue of the Liouvillian it can be obtained for much larger systems (up to $N\approx 1000$ provided the Liouvillian is sparse) using variants of the Lanczos algorithm. This reduces the computational cost of obtaining the NESS, but this reduction is accompanied by a square-root reduction in the sample size which needs to be compensated by more sampling. In other words, the computational advantage of studying the eigenspacing statistics of the NESS lies in being able to explore large system Hilbert space to understand the scaling with $N$.

\begin{figure}[t!]
\begin{center}
\includegraphics[width=\columnwidth]{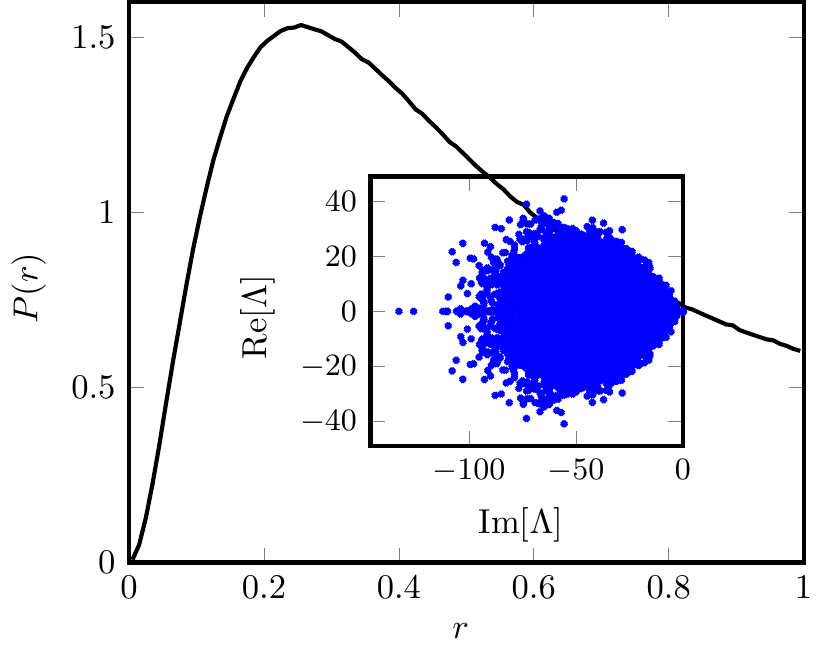}
\end{center}
\caption{The probability distribution $P(r)$ and the inset shows the eigenvalues of the Liouvillian $\Lambda$ confined to the ground-symmetric subspace . The `lemon' shape~\cite{Note2} is distinct near the origin for the eigenvalues of the Liouvillian. The system Hamiltonian is chosen such that $J=1$ and the distribution is obtained over $10^7$ samples. In the inset we plot the eigenvalues only for $2500$ randomly chosen samples. The jump operators have rates $\Gamma_L=1$ and $\Gamma_R=2$.}
\label{fig:eigenspacingstatistics}
\end{figure}

Although, its computationally lucrative to study the eigenspacing statistics of the NESS to uncover universal features, it is highly nontrivial if the Liouvillian is degenerated and the open system symmetries (weak or strong) are unknown. Our approach based on orthonormalization (Sec.~\ref{sec:3B}) is ideally suited for this case. To illustrate this idea, we consider the same para-Benzene ring as before but choose the jump operators [Eq.~\eqref{eq:jump}] extended to all ground-symmetric states [$|i\rangle$ with $i=1,\cdots,4$; see Eq.~\eqref{eq:transformation}] and then randomly picked from the Ginibre unitary ensemble~\cite{Ginibre65}. To simulate a nonequilibrium situation we choose only two jump operators $L_{\mathrm{x}}/\sqrt{\Gamma_{\mathrm{x}}}$ with $\mathrm{x}=1,2$ whose distribution is given by
\ben
P(L_{\mathrm{x}})=\frac{1}{(2\pi)^{N^2}}\exp\left[-\frac{\Tr[ L_{\mathrm{x}}^*L_{\mathrm{x}}]}{2}\right],
\een
with $N=5$ for the case described above. This allows us to ensure that the randomization process is non-pathological~\cite{Can19} and covers the manifold of all jump operators within the ground-symmetric subsector uniformly. Moreover, the full Liouvillian still has a block diagonal structure between ground-symmetric and anti-symmetric subspaces with three steady states. We use then our orthonormalization procedure outlined in Sec.~\ref{sec:3B} and evaluate the distribution of the ratio of consecutive eigenspacing~\cite{Huse07},
\ben
0\leq r_n = \frac{\text{min}\{s_n,s_{n-1}\}}{\text{max}\{s_n,s_{n-1}\}}\leq 1
\een
with $s_n = \nu_{n+1}-\nu_n$ being the eigenspacing of the NESS ($\rho^{\text{NESS}}|\varphi_n\rangle = \nu_n|\varphi_n\rangle$). The ratio, since it is independent of the local density of states, avoids the complications with unfolding of the spectrum and the resulting distribution is  shown in Fig.~\ref{fig:eigenspacingstatistics}. The distribution shows $P(r)\rightarrow 0$ as $r\rightarrow 0$ indicating level repulsion and/or spectral rigidity which means that the NESS is a thermalizing or highly nonintegrable state. The average $\langle r\rangle \sim 0.463$ lies in between the exact predictions from a Poisson and Gaussian orthogonal ensemble (GOE)~\cite{AtlasPRL13}. The inset, Fig.~\ref{fig:eigenspacingstatistics}, shows the distribution of the eigenvalues of the Liouvillian which are available in this case. It should be noted that a more sophisticated form of the sampling could be chosen to obtain a perfect lemon structure~\cite{DenisovPRL19,Note2}, but this does not turn out to be a strict requirement as indicated by the eigenspacing distribution of the the NESS.

Overall, in this section we studied the para-Benzene ring in detail. Although we dealt with the symmetry-decomposition based approach (Sec.~\ref{sec:3A}) throughout this section we would like to end with a few remarks on the other two methods. In all cases, we found that the orthonormalization based approach (Sec.~\ref{sec:3B}) yielded the same results as the symmetry based one. The orthonormalization based approach was also able to treat the ideal-source case and obtain all the six steady states. In complex many-body systems wherein the symmetry operators are either not known or wherein there could be mechanisms due to the baths leading to additional steady states, the orthonormalization based approach is perfectly suited to treat such cases. The large deviation based approach although computationally cheap would fail in the equilibrium and ideal-source situation since the currents for all steady states are zero. This method would also not allow us to obtain the two dark states [Eq.~\eqref{eq:darkstates}] from the anti-symmetric subspace since they both carry zero current. Finally, we ended with studying the eigenspacing distribution of the NESS using the orthonormalization based approach which gave us the expected result that the NESS is a highly non-integrable state. 

\footnotetext[2]{The perfect lemon is achieved when the coherent contribution $-i[H,\rho]$ to the Liouvillian vanishes which is not the case here.}

\section{Conclusions}
\label{sec:5}
In this paper we have presented several techniques to obtain the steady states of a degenerated Lindblad Liouvillian. Each method comes with advantages and disadvantages and, together, they form a useful toolbox for many different problems. First, we have presented a method based on the use of symmetry operators. This technique allows the analytical resolution of many systems, but it requires the existence and knowledge of the open system symmetry operators. The second method is based on a Gramm-Schmidt orthonormalisation is general but computationally expensive. Its utility depends on the degree of degeneracy and on the system dimension. Finally, we have presented a method based on large deviations theory. It does not require any previous knowledge about the system symmetries and it is also computationally cheap as it only requires the diagonalization of an operator of the same size as the Liouvillian. On the other hand, it only gives the density matrices that maximise or minimise a given flux. 

These methods have been illustrated by a canonical example, a para-benzene ring. This system can be analytically diagonalised and in several specific cases it shows a rich phenomenology including dark-states, oscillating coherences, and steady-states that are not a consequence of symmetries. Finally, we have also studied the eigenspacing distribution of the NESS obtained via the orthonormalization method. Since the system by construction is a thermalizing open quantum system the eigenspacing distribution $P(r)\rightarrow 0$ as $r\rightarrow 0$.

There are still several open question to be addressed in this field of research. The para-Benzene ring considered herein had only one NESS, whereas the other steady states were pure. An interesting question remains whether it is possible to construct open quantum systems with more than one NESS, i.e., steady states influenced by the reservoir. Consequently, would these states belong to the same random matrix ensembles, and if they do not, what could be the consequences on observables such as heat and particle currents. Furthermore, the existence of trace zero steady-states has been recently probed but the consequence of these states has not been analysed so far. How do they affect the physical properties of the system and how can they be engineered and detected remains open.

\section*{Acknowledgments}
J.T. acknowledges support by the Institute for Basic Science in Korea (IBS-R024-Y2). D.M.  acknowledges the Spanish Ministry and the Agencia Espa\~nola de Investigaci\'on (AEI) for financial support under grant FIS2017-84256-P (FEDER funds).
%\nocite{*}

\section*{Data Availability}
The data that support the findings of this study are available from the corresponding author upon reasonable request.

%\bibliography{phys.bib}

\begin{thebibliography}{42}%
\makeatletter
\providecommand \@ifxundefined [1]{%
 \@ifx{#1\undefined}
}%
\providecommand \@ifnum [1]{%
 \ifnum #1\expandafter \@firstoftwo
 \else \expandafter \@secondoftwo
 \fi
}%
\providecommand \@ifx [1]{%
 \ifx #1\expandafter \@firstoftwo
 \else \expandafter \@secondoftwo
 \fi
}%
\providecommand \natexlab [1]{#1}%
\providecommand \enquote  [1]{``#1''}%
\providecommand \bibnamefont  [1]{#1}%
\providecommand \bibfnamefont [1]{#1}%
\providecommand \citenamefont [1]{#1}%
\providecommand \href@noop [0]{\@secondoftwo}%
\providecommand \href [0]{\begingroup \@sanitize@url \@href}%
\providecommand \@href[1]{\@@startlink{#1}\@@href}%
\providecommand \@@href[1]{\endgroup#1\@@endlink}%
\providecommand \@sanitize@url [0]{\catcode `\\12\catcode `\$12\catcode
  `\&12\catcode `\#12\catcode `\^12\catcode `\_12\catcode `\%12\relax}%
\providecommand \@@startlink[1]{}%
\providecommand \@@endlink[0]{}%
\providecommand \url  [0]{\begingroup\@sanitize@url \@url }%
\providecommand \@url [1]{\endgroup\@href {#1}{\urlprefix }}%
\providecommand \urlprefix  [0]{URL }%
\providecommand \Eprint [0]{\href }%
\providecommand \doibase [0]{http://dx.doi.org/}%
\providecommand \selectlanguage [0]{\@gobble}%
\providecommand \bibinfo  [0]{\@secondoftwo}%
\providecommand \bibfield  [0]{\@secondoftwo}%
\providecommand \translation [1]{[#1]}%
\providecommand \BibitemOpen [0]{}%
\providecommand \bibitemStop [0]{}%
\providecommand \bibitemNoStop [0]{.\EOS\space}%
\providecommand \EOS [0]{\spacefactor3000\relax}%
\providecommand \BibitemShut  [1]{\csname bibitem#1\endcsname}%
\let\auto@bib@innerbib\@empty
%</preamble>
\bibitem [{\citenamefont {Gorini}, \citenamefont {Kossakowski},\ and\
  \citenamefont {Sudarsahan}(1976)}]{gorini:jmp76}%
  \BibitemOpen
  \bibfield  {author} {\bibinfo {author} {\bibfnamefont {V.}~\bibnamefont
  {Gorini}}, \bibinfo {author} {\bibfnamefont {A.}~\bibnamefont {Kossakowski}},
  \ and\ \bibinfo {author} {\bibfnamefont {E.}~\bibnamefont {Sudarsahan}},\
  }\href@noop {} {\bibfield  {journal} {\bibinfo  {journal} {J. Math. Phys.}\
  }\textbf {\bibinfo {volume} {17}},\ \bibinfo {pages} {821} (\bibinfo {year}
  {1976})}\BibitemShut {NoStop}%
\bibitem [{\citenamefont {Lindblad}(1976)}]{lindblad:cmp76}%
  \BibitemOpen
  \bibfield  {author} {\bibinfo {author} {\bibfnamefont {G.}~\bibnamefont
  {Lindblad}},\ }\href@noop {} {\bibfield  {journal} {\bibinfo  {journal}
  {Commun. Math. Phys.}\ }\textbf {\bibinfo {volume} {119}},\ \bibinfo {pages}
  {48} (\bibinfo {year} {1976})}\BibitemShut {NoStop}%
\bibitem [{\citenamefont {Olmos}, \citenamefont {Lesanovsky},\ and\
  \citenamefont {Garrahan}(2012)}]{olmos:prl12}%
  \BibitemOpen
  \bibfield  {author} {\bibinfo {author} {\bibfnamefont {B.}~\bibnamefont
  {Olmos}}, \bibinfo {author} {\bibfnamefont {I.}~\bibnamefont {Lesanovsky}}, \
  and\ \bibinfo {author} {\bibfnamefont {J.}~\bibnamefont {Garrahan}},\
  }\href@noop {} {\bibfield  {journal} {\bibinfo  {journal} {Phys. Rev. Lett.}\
  }\textbf {\bibinfo {volume} {109}},\ \bibinfo {pages} {020403} (\bibinfo
  {year} {2012})}\BibitemShut {NoStop}%
\bibitem [{\citenamefont {Manzano}\ and\ \citenamefont
  {Kyoseva}(2016)}]{manzano:sr16}%
  \BibitemOpen
  \bibfield  {author} {\bibinfo {author} {\bibfnamefont {D.}~\bibnamefont
  {Manzano}}\ and\ \bibinfo {author} {\bibfnamefont {E.}~\bibnamefont
  {Kyoseva}},\ }\href@noop {} {\bibfield  {journal} {\bibinfo  {journal} {Sci.
  Rep.}\ }\textbf {\bibinfo {volume} {6}},\ \bibinfo {pages} {31161} (\bibinfo
  {year} {2016})}\BibitemShut {NoStop}%
\bibitem [{\citenamefont {Han}\ \emph {et~al.}(2020)\citenamefont {Han},
  \citenamefont {Leykam}, \citenamefont {Angelakis},\ and\ \citenamefont
  {Thingna}}]{Hanarxiv20}%
  \BibitemOpen
  \bibfield  {author} {\bibinfo {author} {\bibfnamefont {J.}~\bibnamefont
  {Han}}, \bibinfo {author} {\bibfnamefont {D.}~\bibnamefont {Leykam}},
  \bibinfo {author} {\bibfnamefont {D.}~\bibnamefont {Angelakis}}, \ and\
  \bibinfo {author} {\bibfnamefont {J.}~\bibnamefont {Thingna}},\ }\href@noop
  {} {\  (\bibinfo {year} {2020})},\ \Eprint {http://arxiv.org/abs/2011.02663}
  {arXiv:2011.02663} \BibitemShut {NoStop}%
\bibitem [{\citenamefont {Thingna}, \citenamefont {Esposito},\ and\
  \citenamefont {Barra}(2019)}]{thingnapre16}%
  \BibitemOpen
  \bibfield  {author} {\bibinfo {author} {\bibfnamefont {J.}~\bibnamefont
  {Thingna}}, \bibinfo {author} {\bibfnamefont {M.}~\bibnamefont {Esposito}}, \
  and\ \bibinfo {author} {\bibfnamefont {F.}~\bibnamefont {Barra}},\
  }\href@noop {} {\bibfield  {journal} {\bibinfo  {journal} {Phys. Rev. E}\
  }\textbf {\bibinfo {volume} {99}},\ \bibinfo {pages} {042142} (\bibinfo
  {year} {2019})}\BibitemShut {NoStop}%
\bibitem [{\citenamefont {Chiara}\ \emph {et~al.}(2018)\citenamefont {Chiara},
  \citenamefont {Landi}, \citenamefont {Hewgill}, \citenamefont {Reid},
  \citenamefont {Ferraro}, \citenamefont {Roncaglia},\ and\ \citenamefont
  {Antezza}}]{chiara:18}%
  \BibitemOpen
  \bibfield  {author} {\bibinfo {author} {\bibfnamefont {G.~D.}\ \bibnamefont
  {Chiara}}, \bibinfo {author} {\bibfnamefont {G.}~\bibnamefont {Landi}},
  \bibinfo {author} {\bibfnamefont {A.}~\bibnamefont {Hewgill}}, \bibinfo
  {author} {\bibfnamefont {B.}~\bibnamefont {Reid}}, \bibinfo {author}
  {\bibfnamefont {A.}~\bibnamefont {Ferraro}}, \bibinfo {author} {\bibfnamefont
  {A.}~\bibnamefont {Roncaglia}}, \ and\ \bibinfo {author} {\bibfnamefont
  {M.}~\bibnamefont {Antezza}},\ }\href@noop {} {\bibfield  {journal} {\bibinfo
   {journal} {New J. Phys.}\ }\textbf {\bibinfo {volume} {20}},\ \bibinfo
  {pages} {113024} (\bibinfo {year} {2018})}\BibitemShut {NoStop}%
\bibitem [{\citenamefont {Liu}, \citenamefont {Segal},\ and\ \citenamefont
  {Hanna}(2019)}]{LiuJPCC19}%
  \BibitemOpen
  \bibfield  {author} {\bibinfo {author} {\bibfnamefont {J.}~\bibnamefont
  {Liu}}, \bibinfo {author} {\bibfnamefont {D.}~\bibnamefont {Segal}}, \ and\
  \bibinfo {author} {\bibfnamefont {G.}~\bibnamefont {Hanna}},\ }\href@noop {}
  {\bibfield  {journal} {\bibinfo  {journal} {J. Phys. Chem. C}\ }\textbf
  {\bibinfo {volume} {123}},\ \bibinfo {pages} {18303} (\bibinfo {year}
  {2019})}\BibitemShut {NoStop}%
\bibitem [{\citenamefont {Quach}\ and\ \citenamefont
  {Munro}(2020)}]{quach:prr20}%
  \BibitemOpen
  \bibfield  {author} {\bibinfo {author} {\bibfnamefont {J.~Q.}\ \bibnamefont
  {Quach}}\ and\ \bibinfo {author} {\bibfnamefont {W.~J.}\ \bibnamefont
  {Munro}},\ }\href@noop {} {\bibfield  {journal} {\bibinfo  {journal} {Phys.
  Rev. App.}\ }\textbf {\bibinfo {volume} {14}},\ \bibinfo {pages} {024092}
  (\bibinfo {year} {2020})}\BibitemShut {NoStop}%
\bibitem [{\citenamefont {Tejero}, \citenamefont {Thingna},\ and\ \citenamefont
  {Manzano}(2020)}]{Manzanoarxiv20}%
  \BibitemOpen
  \bibfield  {author} {\bibinfo {author} {\bibfnamefont {A.}~\bibnamefont
  {Tejero}}, \bibinfo {author} {\bibfnamefont {J.}~\bibnamefont {Thingna}}, \
  and\ \bibinfo {author} {\bibfnamefont {D.}~\bibnamefont {Manzano}},\
  }\href@noop {} {\  (\bibinfo {year} {2020})},\ \Eprint
  {http://arxiv.org/abs/2012.08224} {arXiv:2012.08224} \BibitemShut {NoStop}%
\bibitem [{\citenamefont {\v{Z}nidari\v{c}}, \citenamefont {\v{Z}unkovi\v{c}},\
  and\ \citenamefont {Prosen}(2011)}]{znidaric:pre11}%
  \BibitemOpen
  \bibfield  {author} {\bibinfo {author} {\bibfnamefont {M.}~\bibnamefont
  {\v{Z}nidari\v{c}}}, \bibinfo {author} {\bibfnamefont {B.}~\bibnamefont
  {\v{Z}unkovi\v{c}}}, \ and\ \bibinfo {author} {\bibfnamefont
  {T.}~\bibnamefont {Prosen}},\ }\href@noop {} {\bibfield  {journal} {\bibinfo
  {journal} {Phys. Rev. E}\ }\textbf {\bibinfo {volume} {84}},\ \bibinfo
  {pages} {051115} (\bibinfo {year} {2011})}\BibitemShut {NoStop}%
\bibitem [{\citenamefont {Thingna}, \citenamefont {Garc\'{\i}a-Palacios},\ and\
  \citenamefont {Wang}(2012)}]{thingnaprb12}%
  \BibitemOpen
  \bibfield  {author} {\bibinfo {author} {\bibfnamefont {J.}~\bibnamefont
  {Thingna}}, \bibinfo {author} {\bibfnamefont {J.}~\bibnamefont
  {Garc\'{\i}a-Palacios}}, \ and\ \bibinfo {author} {\bibfnamefont {J.-S.}\
  \bibnamefont {Wang}},\ }\href@noop {} {\bibfield  {journal} {\bibinfo
  {journal} {Phys. Rev. B}\ }\textbf {\bibinfo {volume} {85}},\ \bibinfo
  {pages} {195452} (\bibinfo {year} {2012})}\BibitemShut {NoStop}%
\bibitem [{\citenamefont {Asadian}\ \emph {et~al.}(2013)\citenamefont
  {Asadian}, \citenamefont {Manzano}, \citenamefont {Tiersch},\ and\
  \citenamefont {Briegel}}]{asadian:pre13}%
  \BibitemOpen
  \bibfield  {author} {\bibinfo {author} {\bibfnamefont {A.}~\bibnamefont
  {Asadian}}, \bibinfo {author} {\bibfnamefont {D.}~\bibnamefont {Manzano}},
  \bibinfo {author} {\bibfnamefont {M.}~\bibnamefont {Tiersch}}, \ and\
  \bibinfo {author} {\bibfnamefont {H.}~\bibnamefont {Briegel}},\ }\href@noop
  {} {\bibfield  {journal} {\bibinfo  {journal} {Phys. Rev. E}\ }\textbf
  {\bibinfo {volume} {87}},\ \bibinfo {pages} {012109} (\bibinfo {year}
  {2013})}\BibitemShut {NoStop}%
\bibitem [{\citenamefont {Manzano}, \citenamefont {Chuang},\ and\ \citenamefont
  {Cao}(2016)}]{manzano:njp16}%
  \BibitemOpen
  \bibfield  {author} {\bibinfo {author} {\bibfnamefont {D.}~\bibnamefont
  {Manzano}}, \bibinfo {author} {\bibfnamefont {C.}~\bibnamefont {Chuang}}, \
  and\ \bibinfo {author} {\bibfnamefont {J.}~\bibnamefont {Cao}},\ }\href@noop
  {} {\bibfield  {journal} {\bibinfo  {journal} {New J. Phys.}\ }\textbf
  {\bibinfo {volume} {18}},\ \bibinfo {pages} {043044} (\bibinfo {year}
  {2016})}\BibitemShut {NoStop}%
\bibitem [{\citenamefont {Hu}, \citenamefont {Xia},\ and\ \citenamefont
  {Kais}(2020)}]{hu:sr20}%
  \BibitemOpen
  \bibfield  {author} {\bibinfo {author} {\bibfnamefont {Z.}~\bibnamefont
  {Hu}}, \bibinfo {author} {\bibfnamefont {R.}~\bibnamefont {Xia}}, \ and\
  \bibinfo {author} {\bibfnamefont {S.}~\bibnamefont {Kais}},\ }\href@noop {}
  {\bibfield  {journal} {\bibinfo  {journal} {Sci. Rep.}\ }\textbf {\bibinfo
  {volume} {10}} (\bibinfo {year} {2020})}\BibitemShut {NoStop}%
\bibitem [{\citenamefont {Kraus}\ \emph {et~al.}(2008)\citenamefont {Kraus},
  \citenamefont {B\"uchler}, \citenamefont {Diehl}, \citenamefont {Kantian},
  \citenamefont {Micheli},\ and\ \citenamefont {Zoller}}]{kraus:08}%
  \BibitemOpen
  \bibfield  {author} {\bibinfo {author} {\bibfnamefont {B.}~\bibnamefont
  {Kraus}}, \bibinfo {author} {\bibfnamefont {H.}~\bibnamefont {B\"uchler}},
  \bibinfo {author} {\bibfnamefont {S.}~\bibnamefont {Diehl}}, \bibinfo
  {author} {\bibfnamefont {A.}~\bibnamefont {Kantian}}, \bibinfo {author}
  {\bibfnamefont {A.}~\bibnamefont {Micheli}}, \ and\ \bibinfo {author}
  {\bibfnamefont {P.}~\bibnamefont {Zoller}},\ }\href@noop {} {\bibfield
  {journal} {\bibinfo  {journal} {Phys. Rev. A}\ } (\bibinfo {year}
  {2008})}\BibitemShut {NoStop}%
\bibitem [{\citenamefont {Breuer}\ and\ \citenamefont
  {Petruccione}(2002)}]{breuer_02}%
  \BibitemOpen
  \bibfield  {author} {\bibinfo {author} {\bibfnamefont {H.}~\bibnamefont
  {Breuer}}\ and\ \bibinfo {author} {\bibfnamefont {F.}~\bibnamefont
  {Petruccione}},\ }\href@noop {} {\emph {\bibinfo {title} {The theory of open
  quantum systems}}}\ (\bibinfo  {publisher} {Oxford University Press},\
  \bibinfo {year} {2002})\BibitemShut {NoStop}%
\bibitem [{\citenamefont {Manzano}(2020)}]{manzano:aip20}%
  \BibitemOpen
  \bibfield  {author} {\bibinfo {author} {\bibfnamefont {D.}~\bibnamefont
  {Manzano}},\ }\href@noop {} {\bibfield  {journal} {\bibinfo  {journal} {AIP
  Adv.}\ }\textbf {\bibinfo {volume} {10}},\ \bibinfo {pages} {025106}
  (\bibinfo {year} {2020})}\BibitemShut {NoStop}%
\bibitem [{\citenamefont {Evans}\ and\ \citenamefont
  {Hance-Olsen}(1979)}]{evans:jfa79}%
  \BibitemOpen
  \bibfield  {author} {\bibinfo {author} {\bibfnamefont {D.}~\bibnamefont
  {Evans}}\ and\ \bibinfo {author} {\bibfnamefont {H.}~\bibnamefont
  {Hance-Olsen}},\ }\href@noop {} {\bibfield  {journal} {\bibinfo  {journal}
  {J. Funct. Anal.}\ }\textbf {\bibinfo {volume} {32}},\ \bibinfo {pages} {207}
  (\bibinfo {year} {1979})}\BibitemShut {NoStop}%
\bibitem [{\citenamefont {Bu\v{c}a}\ and\ \citenamefont
  {Prosen}(2012)}]{buca:njp12}%
  \BibitemOpen
  \bibfield  {author} {\bibinfo {author} {\bibfnamefont {B.}~\bibnamefont
  {Bu\v{c}a}}\ and\ \bibinfo {author} {\bibfnamefont {T.}~\bibnamefont
  {Prosen}},\ }\href@noop {} {\bibfield  {journal} {\bibinfo  {journal} {New J.
  Phys.}\ }\textbf {\bibinfo {volume} {14}},\ \bibinfo {pages} {073007}
  (\bibinfo {year} {2012})}\BibitemShut {NoStop}%
\bibitem [{\citenamefont {Manzano}\ and\ \citenamefont
  {Hurtado}(2018)}]{manzano:av18}%
  \BibitemOpen
  \bibfield  {author} {\bibinfo {author} {\bibfnamefont {D.}~\bibnamefont
  {Manzano}}\ and\ \bibinfo {author} {\bibfnamefont {P.}~\bibnamefont
  {Hurtado}},\ }\href@noop {} {\bibfield  {journal} {\bibinfo  {journal} {Adv.
  Phys.}\ }\textbf {\bibinfo {volume} {67}},\ \bibinfo {pages} {1} (\bibinfo
  {year} {2018})}\BibitemShut {NoStop}%
\bibitem [{\citenamefont {Thingna}, \citenamefont {Manzano},\ and\
  \citenamefont {Cao}(2020)}]{thingna:njp20}%
  \BibitemOpen
  \bibfield  {author} {\bibinfo {author} {\bibfnamefont {J.}~\bibnamefont
  {Thingna}}, \bibinfo {author} {\bibfnamefont {D.}~\bibnamefont {Manzano}}, \
  and\ \bibinfo {author} {\bibfnamefont {J.}~\bibnamefont {Cao}},\ }\href@noop
  {} {\bibfield  {journal} {\bibinfo  {journal} {New J. Phys.}\ }\textbf
  {\bibinfo {volume} {22}},\ \bibinfo {pages} {083026} (\bibinfo {year}
  {2020})}\BibitemShut {NoStop}%
\bibitem [{\citenamefont {Lieu}\ \emph {et~al.}(2020)\citenamefont {Lieu},
  \citenamefont {Belyansky}, \citenamefont {Young}, \citenamefont {Lundgren},
  \citenamefont {Albert},\ and\ \citenamefont {Gorshkov}}]{lieu:arxiv20}%
  \BibitemOpen
  \bibfield  {author} {\bibinfo {author} {\bibfnamefont {S.}~\bibnamefont
  {Lieu}}, \bibinfo {author} {\bibfnamefont {R.}~\bibnamefont {Belyansky}},
  \bibinfo {author} {\bibfnamefont {J.}~\bibnamefont {Young}}, \bibinfo
  {author} {\bibfnamefont {R.}~\bibnamefont {Lundgren}}, \bibinfo {author}
  {\bibfnamefont {V.}~\bibnamefont {Albert}}, \ and\ \bibinfo {author}
  {\bibfnamefont {A.}~\bibnamefont {Gorshkov}},\ }\href@noop {} {\  (\bibinfo
  {year} {2020})},\ \Eprint {http://arxiv.org/abs/2008.02816}
  {arXiv:2008.02816} \BibitemShut {NoStop}%
\bibitem [{\citenamefont {Fiorelli}\ \emph {et~al.}(2019)\citenamefont
  {Fiorelli}, \citenamefont {Rotondo}, \citenamefont {Marcuzzi}, \citenamefont
  {Garrahan},\ and\ \citenamefont {Lesanovsky}}]{fiorelli:pra19}%
  \BibitemOpen
  \bibfield  {author} {\bibinfo {author} {\bibfnamefont {E.}~\bibnamefont
  {Fiorelli}}, \bibinfo {author} {\bibfnamefont {P.}~\bibnamefont {Rotondo}},
  \bibinfo {author} {\bibfnamefont {M.}~\bibnamefont {Marcuzzi}}, \bibinfo
  {author} {\bibfnamefont {J.}~\bibnamefont {Garrahan}}, \ and\ \bibinfo
  {author} {\bibfnamefont {I.}~\bibnamefont {Lesanovsky}},\ }\href@noop {}
  {\bibfield  {journal} {\bibinfo  {journal} {Phys. Rev. A}\ }\textbf {\bibinfo
  {volume} {99}},\ \bibinfo {pages} {032126} (\bibinfo {year}
  {2019})}\BibitemShut {NoStop}%
\bibitem [{\citenamefont {Thingna}, \citenamefont {Manzano},\ and\
  \citenamefont {Cao}(2016)}]{thingna:sr16}%
  \BibitemOpen
  \bibfield  {author} {\bibinfo {author} {\bibfnamefont {J.}~\bibnamefont
  {Thingna}}, \bibinfo {author} {\bibfnamefont {D.}~\bibnamefont {Manzano}}, \
  and\ \bibinfo {author} {\bibfnamefont {J.}~\bibnamefont {Cao}},\ }\href@noop
  {} {\bibfield  {journal} {\bibinfo  {journal} {Sci. Rep.}\ }\textbf {\bibinfo
  {volume} {6}},\ \bibinfo {pages} {28027} (\bibinfo {year}
  {2016})}\BibitemShut {NoStop}%
\bibitem [{\citenamefont {Prosen}\ and\ \citenamefont {\ifmmode \check{Z}\else
  \v{Z}\fi{}nidari\ifmmode~\check{c}\else \v{c}\fi{}}(2013)}]{ProsenPRL13}%
  \BibitemOpen
  \bibfield  {author} {\bibinfo {author} {\bibfnamefont {T.}~\bibnamefont
  {Prosen}}\ and\ \bibinfo {author} {\bibfnamefont {M.}~\bibnamefont {\ifmmode
  \check{Z}\else \v{Z}\fi{}nidari\ifmmode~\check{c}\else \v{c}\fi{}}},\
  }\href@noop {} {\bibfield  {journal} {\bibinfo  {journal} {Phys. Rev. Lett.}\
  }\textbf {\bibinfo {volume} {111}},\ \bibinfo {pages} {124101} (\bibinfo
  {year} {2013})}\BibitemShut {NoStop}%
\bibitem [{\citenamefont {Albert}\ and\ \citenamefont
  {Jiang}(2014)}]{albert:pra14}%
  \BibitemOpen
  \bibfield  {author} {\bibinfo {author} {\bibfnamefont {V.}~\bibnamefont
  {Albert}}\ and\ \bibinfo {author} {\bibfnamefont {L.}~\bibnamefont {Jiang}},\
  }\href@noop {} {\bibfield  {journal} {\bibinfo  {journal} {Phys. Rev. A}\
  }\textbf {\bibinfo {volume} {89}},\ \bibinfo {pages} {022118} (\bibinfo
  {year} {2014})}\BibitemShut {NoStop}%
\bibitem [{\citenamefont {Prosen}(2012)}]{prosen:ps12}%
  \BibitemOpen
  \bibfield  {author} {\bibinfo {author} {\bibfnamefont {T.}~\bibnamefont
  {Prosen}},\ }\href@noop {} {\bibfield  {journal} {\bibinfo  {journal} {Phys.
  Scr.}\ }\textbf {\bibinfo {volume} {86}},\ \bibinfo {pages} {058511}
  (\bibinfo {year} {2012})}\BibitemShut {NoStop}%
\bibitem [{Note1()}]{Note1}%
  \BibitemOpen
  \bibinfo {note} {Note that duality of basis ensures the left and right
  eigenvectors are form an orthonormal set. This does not ensure that the right
  eigenvectors are orthogonal amongst themselves.}\BibitemShut {Stop}%
\bibitem [{\citenamefont {Zhang}\ \emph {et~al.}(2020)\citenamefont {Zhang},
  \citenamefont {Tindall}, \citenamefont {Mur-Petit}, \citenamefont {Jaksch},\
  and\ \citenamefont {Buca}}]{zhang_jpa20}%
  \BibitemOpen
  \bibfield  {author} {\bibinfo {author} {\bibfnamefont {Z.}~\bibnamefont
  {Zhang}}, \bibinfo {author} {\bibfnamefont {J.}~\bibnamefont {Tindall}},
  \bibinfo {author} {\bibfnamefont {J.}~\bibnamefont {Mur-Petit}}, \bibinfo
  {author} {\bibfnamefont {D.}~\bibnamefont {Jaksch}}, \ and\ \bibinfo {author}
  {\bibfnamefont {B.}~\bibnamefont {Buca}},\ }\href@noop {} {\bibfield
  {journal} {\bibinfo  {journal} {J. Phys. A: Math. Theor.}\ }\textbf {\bibinfo
  {volume} {53}},\ \bibinfo {pages} {215304} (\bibinfo {year}
  {2020})}\BibitemShut {NoStop}%
\bibitem [{\citenamefont {Manzano}\ and\ \citenamefont
  {Hurtado}(2014)}]{manzano:prb14}%
  \BibitemOpen
  \bibfield  {author} {\bibinfo {author} {\bibfnamefont {D.}~\bibnamefont
  {Manzano}}\ and\ \bibinfo {author} {\bibfnamefont {P.}~\bibnamefont
  {Hurtado}},\ }\href@noop {} {\bibfield  {journal} {\bibinfo  {journal} {Phys.
  Rev. B}\ }\textbf {\bibinfo {volume} {90}},\ \bibinfo {pages} {125138}
  (\bibinfo {year} {2014})}\BibitemShut {NoStop}%
\bibitem [{\citenamefont {Denisov}\ \emph {et~al.}(2019)\citenamefont
  {Denisov}, \citenamefont {Laptyeva}, \citenamefont {Tarnowski}, \citenamefont
  {Chru\ifmmode \acute{s}\else \'{s}\fi{}ci\ifmmode~\acute{n}\else
  \'{n}\fi{}ski},\ and\ \citenamefont {\ifmmode~\dot{Z}\else
  \.{Z}\fi{}yczkowski}}]{DenisovPRL19}%
  \BibitemOpen
  \bibfield  {author} {\bibinfo {author} {\bibfnamefont {S.}~\bibnamefont
  {Denisov}}, \bibinfo {author} {\bibfnamefont {T.}~\bibnamefont {Laptyeva}},
  \bibinfo {author} {\bibfnamefont {W.}~\bibnamefont {Tarnowski}}, \bibinfo
  {author} {\bibfnamefont {D.}~\bibnamefont {Chru\ifmmode \acute{s}\else
  \'{s}\fi{}ci\ifmmode~\acute{n}\else \'{n}\fi{}ski}}, \ and\ \bibinfo {author}
  {\bibfnamefont {K.}~\bibnamefont {\ifmmode~\dot{Z}\else
  \.{Z}\fi{}yczkowski}},\ }\href@noop {} {\bibfield  {journal} {\bibinfo
  {journal} {Phys. Rev. Lett.}\ }\textbf {\bibinfo {volume} {123}},\ \bibinfo
  {pages} {140403} (\bibinfo {year} {2019})}\BibitemShut {NoStop}%
\bibitem [{\citenamefont {Wang}, \citenamefont {Piazza},\ and\ \citenamefont
  {Luitz}(2020)}]{DavidePRL20}%
  \BibitemOpen
  \bibfield  {author} {\bibinfo {author} {\bibfnamefont {K.}~\bibnamefont
  {Wang}}, \bibinfo {author} {\bibfnamefont {F.}~\bibnamefont {Piazza}}, \ and\
  \bibinfo {author} {\bibfnamefont {D.~J.}\ \bibnamefont {Luitz}},\ }\href@noop
  {} {\bibfield  {journal} {\bibinfo  {journal} {Phys. Rev. Lett.}\ }\textbf
  {\bibinfo {volume} {124}},\ \bibinfo {pages} {100604} (\bibinfo {year}
  {2020})}\BibitemShut {NoStop}%
\bibitem [{\citenamefont {S\'a}, \citenamefont {Ribeiro},\ and\ \citenamefont
  {Prosen}(2020)}]{ProsenPRX20}%
  \BibitemOpen
  \bibfield  {author} {\bibinfo {author} {\bibfnamefont {L.}~\bibnamefont
  {S\'a}}, \bibinfo {author} {\bibfnamefont {P.}~\bibnamefont {Ribeiro}}, \
  and\ \bibinfo {author} {\bibfnamefont {T.}~\bibnamefont {Prosen}},\
  }\href@noop {} {\bibfield  {journal} {\bibinfo  {journal} {Phys. Rev. X}\
  }\textbf {\bibinfo {volume} {10}},\ \bibinfo {pages} {021019} (\bibinfo
  {year} {2020})}\BibitemShut {NoStop}%
\bibitem [{\citenamefont {Berry}\ and\ \citenamefont {Tabor}(1977)}]{Berry77}%
  \BibitemOpen
  \bibfield  {author} {\bibinfo {author} {\bibfnamefont {M.}~\bibnamefont
  {Berry}}\ and\ \bibinfo {author} {\bibfnamefont {M.}~\bibnamefont {Tabor}},\
  }\href@noop {} {\bibfield  {journal} {\bibinfo  {journal} {Proc. R. Soc.
  Lond. A}\ }\textbf {\bibinfo {volume} {356}},\ \bibinfo {pages} {375}
  (\bibinfo {year} {1977})}\BibitemShut {NoStop}%
\bibitem [{\citenamefont {Berry}(1981)}]{BerryAP81}%
  \BibitemOpen
  \bibfield  {author} {\bibinfo {author} {\bibfnamefont {M.}~\bibnamefont
  {Berry}},\ }\href@noop {} {\bibfield  {journal} {\bibinfo  {journal} {Ann.
  Phys.}\ }\textbf {\bibinfo {volume} {131}},\ \bibinfo {pages} {163 }
  (\bibinfo {year} {1981})}\BibitemShut {NoStop}%
\bibitem [{\citenamefont {Mehta}(2004)}]{Mehta}%
  \BibitemOpen
  \bibfield  {author} {\bibinfo {author} {\bibfnamefont {M.}~\bibnamefont
  {Mehta}},\ }\href@noop {} {\emph {\bibinfo {title} {Random Matrices}}}\
  (\bibinfo  {publisher} {Elsevier, New York},\ \bibinfo {year}
  {2004})\BibitemShut {NoStop}%
\bibitem [{\citenamefont {Ginibre}(1965)}]{Ginibre65}%
  \BibitemOpen
  \bibfield  {author} {\bibinfo {author} {\bibfnamefont {J.}~\bibnamefont
  {Ginibre}},\ }\href@noop {} {\bibfield  {journal} {\bibinfo  {journal} {J.
  Math. Phys.}\ }\textbf {\bibinfo {volume} {6}},\ \bibinfo {pages} {440}
  (\bibinfo {year} {1965})}\BibitemShut {NoStop}%
\bibitem [{\citenamefont {Can}(2019)}]{Can19}%
  \BibitemOpen
  \bibfield  {author} {\bibinfo {author} {\bibfnamefont {T.}~\bibnamefont
  {Can}},\ }\href@noop {} {\bibfield  {journal} {\bibinfo  {journal} {J. Phys.
  A: Math. Theor.}\ }\textbf {\bibinfo {volume} {52}},\ \bibinfo {pages}
  {485302} (\bibinfo {year} {2019})},\ \bibinfo {note} {see simple dissipator
  herein.}\BibitemShut {Stop}%
\bibitem [{\citenamefont {Oganesyan}\ and\ \citenamefont
  {Huse}(2007)}]{Huse07}%
  \BibitemOpen
  \bibfield  {author} {\bibinfo {author} {\bibfnamefont {V.}~\bibnamefont
  {Oganesyan}}\ and\ \bibinfo {author} {\bibfnamefont {D.~A.}\ \bibnamefont
  {Huse}},\ }\href {\doibase 10.1103/PhysRevB.75.155111} {\bibfield  {journal}
  {\bibinfo  {journal} {Phys. Rev. B}\ }\textbf {\bibinfo {volume} {75}},\
  \bibinfo {pages} {155111} (\bibinfo {year} {2007})}\BibitemShut {NoStop}%
\bibitem [{\citenamefont {Atas}\ \emph {et~al.}(2013)\citenamefont {Atas},
  \citenamefont {Bogomolny}, \citenamefont {Giraud},\ and\ \citenamefont
  {Roux}}]{AtlasPRL13}%
  \BibitemOpen
  \bibfield  {author} {\bibinfo {author} {\bibfnamefont {Y.~Y.}\ \bibnamefont
  {Atas}}, \bibinfo {author} {\bibfnamefont {E.}~\bibnamefont {Bogomolny}},
  \bibinfo {author} {\bibfnamefont {O.}~\bibnamefont {Giraud}}, \ and\ \bibinfo
  {author} {\bibfnamefont {G.}~\bibnamefont {Roux}},\ }\href {\doibase
  10.1103/PhysRevLett.110.084101} {\bibfield  {journal} {\bibinfo  {journal}
  {Phys. Rev. Lett.}\ }\textbf {\bibinfo {volume} {110}},\ \bibinfo {pages}
  {084101} (\bibinfo {year} {2013})}\BibitemShut {NoStop}%
\bibitem [{Note2()}]{Note2}%
  \BibitemOpen
  \bibinfo {note} {The perfect lemon is achieved when the coherent contribution
  $-i[H,\rho ]$ to the Liouvillian vanishes which is not the case
  here.}\BibitemShut {Stop}%
\end{thebibliography}
\bibliographystyle{aipnum4-1.bst}
\end{document}